\pdfoutput=1
\documentclass[reprint,superscriptaddress,nofootinbib,amsmath,amssymb,aps,prd,noeprint,twocolumn]{revtex4-2} 

\usepackage{graphicx}
\usepackage{dcolumn}
\usepackage{bm}
\usepackage{xcolor}
\usepackage{hyperref}
\hypersetup{
  colorlinks=true,
  linkcolor=blue,                
  citecolor=green!50!black,      
  urlcolor=magenta,              
}
\usepackage{orcidlink}
\usepackage{booktabs}
\usepackage{makecell}
\usepackage{multirow}

\makeatletter
\newcount\bib@nauth
\newtoks\bib@authtoks
\def\bib@authorstr{author}
\def\bib@collect#1#2{
  \global\advance\bib@nauth\@ne
  \ifnum\bib@nauth<4
    \global\bib@authtoks\expandafter{\the\bib@authtoks\bibinfo{author}{#2}}%
    \ifnum\bib@nauth<3 \global\bib@authtoks\expandafter{\the\bib@authtoks,\ }\fi
  \fi}
\def\bib@newfield#1#2{%
  \def\bib@tmp{#1}%
  \ifx\bib@tmp\bib@authorstr
    \global\bib@nauth\z@ \global\bib@authtoks{}%
    \setbox\z@\hbox{\let\bibinfo\bib@collect #2}%
    \ifnum\bib@nauth>3
      \global\bib@authtoks\expandafter{\the\bib@authtoks, \emph{et~al.},\ }%
      \expandafter\bib@authorfield\expandafter{\the\bib@authtoks}%
    \else\bib@origfield{#1}{#2}\fi
  \else\bib@origfield{#1}{#2}\fi}
\def\bib@authorfield{\bib@origfield{author}}
\AtBeginDocument{\let\bib@origfield\bibfield \let\bibfield\bib@newfield}
\makeatother

\newcommand{\eqn}[1]{eq.~(#1)}
\newcommand{\fg}[1]{Fig.~#1}
\newcommand{\tab}[1]{Table~#1}
\newcommand{\mrm}[1]{\mathrm{#1}}
\newcommand{\dd}{\mathrm{d}}
\newcommand{\expnumber}[2]{{#1}\mathrm{e}{#2}}
\newcommand{\Ndist}[3]{ \mathcal{N}({#1},\,\expnumber{{#2}}{{#3}}) } 
\newcommand{\Udist}[2]{ \mathcal{U}({#1},\,{#2}) } 

\newcommand{\CL}{\textsc{CosmoLike}\ }
\newcommand{\lcdm}{\(\Lambda\mrm{CDM}\)}
\newcommand{\wcdm}{\(w_0 w_a\mrm{CDM}\)}

\begin{document}

\title{Multiprobe strategies for Kinematic Lensing with the Roman Space Telescope High Latitude Wide Area Survey}

\author{Yu-Hsiu Huang\orcidlink{0000-0002-4982-0208}} 
\affiliation{Department of Astronomy, University of Arizona, Tucson, AZ 85721, USA}

\author{Jiachuan Xu\orcidlink{0000-0003-0871-8941}}
\affiliation{Department of Physics, Northeastern University, Boston, MA, 02115, USA}
\affiliation{Department of Astronomy, University of Arizona, Tucson, AZ 85721, USA}

\author{Tim Eifler\orcidlink{0000-0002-1894-3301}}
\affiliation{Department of Astronomy, University of Arizona, Tucson, AZ 85721, USA}
\affiliation{Department of Physics, University of Arizona, Tucson, AZ 85721, USA}

\author{Elisabeth Krause\orcidlink{0000-0001-8356-2014}}%
\affiliation{Department of Astronomy, University of Arizona, Tucson, AZ 85721, USA}
\affiliation{Department of Physics, University of Arizona, Tucson, AZ 85721, USA}

\author{Spencer Everett\orcidlink{0000-0002-3745-2882}}
\affiliation{California Institute of Technology, Pasadena, California 91109, USA}

\author{Fabian Hervas-Peters\orcidlink{0009-0008-1839-2969}}
\affiliation{Department of Astronomy, University of Arizona, Tucson, AZ 85721, USA}

\author{Pranjal R.~S.\orcidlink{0000-0003-3714-2574}}
\affiliation{Jet Propulsion Laboratory, California Institute of Technology, Pasadena, California 91109, USA}

\author{Eric Huff\orcidlink{0000-0002-9378-3424}}
\affiliation{Jet Propulsion Laboratory, California Institute of Technology, Pasadena, California 91109, USA}

\author{Yan Xiang Lai\orcidlink{s0000-0001-9054-4324}}
\affiliation{Department of Astronomy, University of Arizona, Tucson, AZ 85721, USA}

\date{\today}

\begin{abstract}
In its latest design, the High Latitude Wide Area Survey (HLWAS) Medium Tier comprises overlapping photometric and spectroscopic data across 2,300 square degrees. This dataset enables Kinematic Lensing (KL), a Weak Lensing (WL) technique that exploits galaxy kinematics to control for shape noise, the dominant statistical limitation of traditional WL. We present several simulated analyses quantifying the expected cosmological science return of a Roman KL analysis with the latest design of the HLWAS: We explore KL prospects with early Roman data from the Deep Tier fields and from the early Medium Tier fields; we demonstrate the impact of different roll-angle choices on the KL performance; we benchmark KL against traditional WL; we introduce a KL-based multiprobe analysis that jointly models KL, galaxy clustering, and galaxy-galaxy lensing. In these analyses, we consider systematics, including photometric uncertainties, shear calibration bias, galaxy bias, and baryonic feedback. Our results show that KL alone achieves a dark energy figure of merit (FoM) of 22.54, which is more than a factor of two improvement over a corresponding WL analysis. The KL multiprobe approach improves the full-survey FoM to 34.58, even assuming conservative scale cuts in the clustering and galaxy-galaxy lensing. For the HLWAS roll-angle strategy, we find that the current plan of 4 passes with rotations of approximately 0, 10, 170, 180 degrees are near optimal for KL.  

\end{abstract}


\maketitle
\section{Introduction}
Dark energy accounts for roughly 70\% of the mass and energy in the universe \citep{Planck2018, DESY5_SN}. Its physical origin, however, remains unexplained. In the absence of a fundamental theory, the standard cosmological model \lcdm{} adopts the simplest phenomenological assumption: a cosmological constant with a fixed equation of state of \(w=-1\). This standard model has provided a statistically good fit to observations spanning the Cosmic Microwave Background (CMB), Baryon Acoustic Oscillation (BAO), and large-scale structure (LSS) individually over decades \citep{Planck2018, SDSS_BAO, KiDS2021:3x2, DES2026:Y6_3x2}.

Recent results from the Dark Energy Spectroscopic Instrument (DESI\footnote{\url{https://data.desi.lbl.gov/doc/}}) challenge the \lcdm{} model. While DESI's BAO data alone remain consistent with \lcdm, the joint analysis of BAO, CMB, and supernova (SN) favors a time-varying equation of state, \(w(a)=w_0 + w_a (1-a)\) (the Chevallier-Polarski-Linder or CPL parameterization; \citep{Chevallier_Polarski_2001_CPL, Linder_2003_CPL}), deviating from the fixed \(w=-1\) assumed in \lcdm{} \citep{Lodha2025_DDE, DESI_DR2_BAO, DESI_DR1_BAO}. Since the release of DESI DR1, this result has been the subject of active debate \citep{Corte_Liddle_2024, Corte_Liddle2025_noSN, Popovic2026_SNsys, Efstathiou2025_SNsys, Liu2024_LRG, Colgain2026, Toomey2026_prior, Efstathiou2025_degeneracy, Herold_Karwal_2026_frequentist, Vincenzi2025_SN_DESI, Adame2025_full_shape}. 

Fully understanding the physics of dark energy and stress-testing the \lcdm{} model requires additional independent measurements. For instance, the Euclid mission\footnote{\url{https://www.cosmos.esa.int/web/euclid}} is designed to measure BAO over similar redshift ranges as DESI, with distinctly different observational systematics \citep{Euclid2024_overview}. Cross-checking DESI's and Euclid's BAO measurements will reveal whether unknown pipeline systematics exist. Similarly, the Nancy Grace Roman Space Telescope (Roman\footnote{\url{https://science.nasa.gov/mission/roman-space-telescope/}}) and Rubin Observatory's Legacy Survey of Space and Time (LSST\footnote{\url{https://www.lsst.org}}) are expected to observe thousands of Type Ia SNe in the coming years \citep{Kessler2025_RomanSN, Mitra2026_LSSTSN}, enabling studies to characterize the astrophysical systematics of SN light curves.

Complementary measurements offer another avenue to probe dark energy physics. Cosmic shear, Weak Lensing (WL) by the large-scale structure, probes the dark energy equation of state through lensing geometry and angular projection, providing information complementary to BAO's standard-ruler measurement. Moreover, it provides additional constraints on structure growth and thus enables a growth-geometry-consistency test of the cosmological model. However, cosmic shear alone is subject to degeneracies between cosmology and nuisance parameters such as intrinsic alignment or photometric redshift uncertainties \citep{Anbajagane2025_DECADE, Secco2022_DESY3_cosmic_shear, Amon2022_DESY3_cosmic_shear}. A typical way to improve cosmic shear constraints is to combine it with galaxy clustering and galaxy-galaxy lensing into a so-called $3\times2$pt analysis. This concept has been applied to the Stage-III photometric surveys such as the Dark Energy Survey (DES\footnote{\url{https://www.darkenergysurvey.org}}), the Kilo-Degree Survey (KiDS\footnote{\url{https://kids.strw.leidenuniv.nl}}), and the Hyper Suprime-Cam Subaru Strategic Program (HSC\footnote{\url{https://hsc-release.mtk.nao.ac.jp/doc/}}) \citep{DES2026:Y6_3x2, KiDS2021:3x2, Sugiyama2023:HSCY3_3x2, Miyatake2023:HSCY3_3x2} to measure dark energy equation of state \citep{DES2026:DE, KiDS2026:DE}. 

Another way to improve cosmic shear constraints is Kinematic Lensing (KL). KL combines spectroscopic and imaging data with the Tully-Fisher relation \citep{TFR} to extract the lensing signal at a much lower shape-noise level than the traditional WL \citep{Huff2013_KL, Xu2023, RS2023_shape_noise}. Beyond reducing shape noise, KL is also immune to several WL systematics such as photometric redshift uncertainties and intrinsic alignment \citep{TED}. The KL concept has been demonstrated in cluster lensing using slit spectroscopy, achieving a shape-noise uncertainty of 0.026 per component \citep{RS2025_pilot}. Moreover, forecasts for several upcoming surveys, particularly Roman, suggest that KL has substantial potential to constrain cosmology \citep{Xu2023, Xu2024, Huang2026, Li2026_KL_lensing_ratio}. 

One of Roman's core surveys is the High Latitude Wide Area Survey (HLWAS), which has four tiers: Wide, Medium, Deep, and Ultra-Deep Tier, covering different survey depths and areas. Among the four tiers, the Medium Tier comprises images and slitless grism spectroscopy over 2,300~deg\(^2\) \citep{ROTAC2025}, providing the essential ingredients for KL measurements. Previously, \citep{Xu2023} showed that KL could improve the constraining power on the dark energy equation of state by more than three times compared to WL, using the now-outdated Medium Tier survey configuration \citep{Eifler2021_Roman_multiprobe}. The recent Roman Observations Time Allocation Committee (ROTAC) decided on a shallower but wider HLWAS Medium Tier \citep{ROTAC2025}, which we forecast and explore in detail in this paper.

Further, we examine the impact of roll-angle separation between different grism exposures. This is motivated by findings in ~\citep{wang_hlss_2022}, which revealed a non-trivial relation between the roll-angle separation and the grism spectra success rate. Small roll angle separations increase the difficulty of spectra decontamination, while large roll angles lead to more positions with poor exposure coverage due to more complex tiling, which we should account for when forecasting KL's performance.

In addition to an in-depth study of KL performance with the new survey design, we explore the multiprobe concepts centered around KL. An intuitive approach is to combine KL with 3D galaxy clustering because the KL sample naturally requires spectroscopic measurements. This cross-correlation also yields the greatest benefit when using overlapping data, such as the Medium Tier of HLWAS \citep{Bernstein2011_overlap, Cai2012_overlap}. However, developing the full formalism including non-Gaussian covariance matrices for a 3D clustering plus KL analysis is beyond the scope of this paper. As a first step, we explore the information gain of a standard 2D tomographic \(3\times2\)pt method with KL as the lensing measurement. We denote this multiprobe approach as KL-\(3\times2\)pt.

The paper is structured as follows. In Sect.~\ref{sec:theory} we introduce our model for the simulated analyses. In Sect.~\ref {sec:sample}, we describe our galaxy sample selection for both KL and WL. In Sect.~\ref {sec:roll}, we explore the relation between the telescope roll angle and the resulting KL shape noise. In Sect.~\ref{sec:forecast} we present and compare the updated KL and WL forecasts for HLWAS. In Sect.~\ref{sec:multiprobe} we present the KL-multiprobe forecast. We conclude the paper in Sect.~\ref{sec:conclusion}.

\section{Theoretical Modeling of observables\label{sec:theory}}

We use \CL\citep{cosmolike} routines as described in \citep{Xu2023} for all lensing-based observables and include the \CL{}FKEM module \citep{Fang2020} to compute non-Limber power spectra for galaxy clustering. We integrate our KL-\CL{} framework into \textsc{Cobaya-CosmoLike Architecture} (\textsc{CoCoA}\footnote{\href{https://github.com/CosmoLike/cocoa}{https://github.com/CosmoLike/cocoa}}) and make the code available to the public\footnote{\url{https://github.com/CosmoLike/cocoa_roman_kl}}.

In all cases, we compute the posterior probability distribution of cosmological parameters $\bm{p}_{\mrm{co}}$ and nuisance parameters $\bm{p}_{\mrm{nu}}$ given the data $\mathbf{D}$ via Bayes' theorem 
\begin{equation}
\label{eqn:post}
\begin{split}
P(\bm{p}|\bm{D})&\propto L(\bm{D}|\bm{p}) \, \pi(\bm{p})\\
    &\propto\mrm{\exp}\left(-\frac{1}{2} \chi^2 (\bm{p})\right) \, \pi(\bm{p}),
\end{split}
\end{equation}
where $\bm{p}=(\bm{p}_{\mrm{co}}, \bm{p}_{\mrm{nu}})$ and $\pi(\bm{p})$ is the prior probability distribution of the parameters.
We further assume that the likelihood $L(\mathbf{D}|\bm{p})$ in \eqn{\ref{eqn:post}} is well-approximated by a multivariate Gaussian and define
\begin{equation}
\begin{aligned}
    \chi^2(\bm{p})=(\mathbf{D}-\mathbf{M}(\bm{p}))^{\rm{T}}\mathbf{Cov}^{-1}(\mathbf{D}-\mathbf{M}(\bm{p})),  
\end{aligned}
\end{equation}
where $\mathbf{M}$ is the model vector evaluated at $\bm{p}$ and $\mathbf{Cov}$ the covariance matrix calculated at the fiducial parameter values. We assume the covariance is constant in parameter space; we do not adopt a fully cosmology-dependent covariance approach in this work \citep{Eifler2009_cos_dep_cov, Kodwani2019_cos_dep_cov}.

\subsection{Data Vector\label{sec:dv}}
We compute all data vectors in Fourier space. For each probe, cosmic shear, galaxy clustering, and galaxy-galaxy lensing, we compute the angular power spectra as 
\begin{equation} \label{eq:nonlimber}
    \begin{split}
        \bm{C}^{ij}_{ab}(\ell)&=\frac{2}{\pi}\int d\chi_1 q^i_a(\chi_1)\int d\chi_2 q^{j}_b(\chi_2)\\
            \int &\frac{dk}{k}k^3P(k,z(\chi_1), z(\chi_2)) j_\ell^{(a)}(k\chi_1) j_\ell^{(b)}(k\chi_2).
    \end{split}
\end{equation}
where $\chi$ is the comoving distance, $q^i_a(\chi)$ is the kernel function at the $i$-th tomographic bin for the tracer $a$, $P(k,z(\chi_1),z(\chi_2))$ is the 3D matter power spectrum, and $j_\ell^{(a)}(k\chi)$ is the spherical Bessel function with the derivative order associated with the probe. When the radial width of the kernel function is larger than $1/k$ of the relevant mode of the tracer, we apply the Limber approximation \citep{LoVerdeAfshordi2008}, where $j_\ell(k\chi)\simeq \delta_D(\ell+1/2-k\chi)$ and  \eqn{\ref{eq:nonlimber}} reduces to 
\begin{equation} \label{eq:limber}
    \bm{C}^{ij}_{ab}(\ell) = \frac{2}{2\ell+1}\int dk\,q^i_a(\chi_\ell) q^j_b(\chi_\ell) P(k,\chi_\ell)
\end{equation}
with $\chi_\ell=(\ell+1/2)/k$. The Limber approximation works well when \(q_a^i(\chi)\) is smooth over a radial width much larger than the scale of tracer \(a\)'s fluctuation. This generally holds for the lensing kernel, which is usually wide. However, when the tomographic bin becomes narrower than the scales of the density fluctuation mode, the approximation breaks down. 

For the forecasting purpose of this paper, where data vectors are generated with the same framework as model vectors, we adopt the Limber approximation as a sufficient modeling choice. We study the impact of this approximation in the narrow tomographic-bin scenario in Sect.~\ref{sec:NL}. We note that while Limber is sufficient for forecasting (error bars between Limber and non-Limber are largely the same), non-Limber is required for an actual data analysis to avoid parameter biases. 

\paragraph{Lensing kernels}
The lensing kernel for cosmic shear at the $i$-th bin is
\begin{equation}
    \begin{split}
        q_\kappa^{i}(\chi)&=\frac{3H_0^2\Omega_\mrm{m}}{2c^2}\frac{\chi}{a(\chi)}\\
        &\int_\chi^{\chi_h}\dd \chi'\frac{n_\mrm{src}^{i}(z(\chi'))}{\bar{n}_\mrm{src}^{i}}\frac{\dd z}{\dd \chi'}\frac{\chi'-\chi}{\chi'}\,,
    \end{split}
\end{equation}
where $n_\mrm{src}^{i}(z)$ represents the redshift distribution of source galaxies in the $i$-th bin, with $\bar{n}_\mrm{src}^{i}$ the 2D angular galaxy density within the same redshift bin.

\paragraph{Clustering kernels}

The selection function for clustering is 
\begin{equation}
    q^i_{g}(\chi)=\frac{n^i_{\rm lens}(z(\chi))}{\bar{n}^i_{\rm lens}} \frac{dz}{d\chi}
\end{equation}
where $n^i_{\rm lens}(z)$ is the redshift distribution of lens galaxies in the $i$-th tomographic bin and $\bar{n}_\mrm{lens}^i$ is the 2D angular galaxy density within the same redshift bin.

\subsection{Covariance Matrix\label{sec:cov}}
The covariance between the two angular power spectra \(\bm{C}_{ab}^{ij}(\ell_1)\) and \(\bm{C}_{cd}^{kl}(\ell_2)\) 
\begin{equation}
\begin{aligned}
    \mathbf{Cov}\left({\bm{C}}_{ab}^{ij}(\ell_1), {\bm{C}}_{cd}^{kl}(\ell_2)\right) &=  \mathbf{Cov}^\mrm{G}\left({\bm{C}}_{ab}^{ij}(\ell_1), {\bm{C}}_{cd}^{kl}(\ell_2)\right)\\
+&\mathbf{Cov}^\mrm{SSC}\left(\bm{C}_{ab}^{ij}(\ell_1), \bm{C}_{cd}^{kl}(\ell_2)\right) \\
+&\mathbf{Cov}^\mrm{cNG}\left(\bm{C}_{ab}^{ij}(\ell_1), \bm{C}_{cd}^{kl}(\ell_2)\right)
\end{aligned}
\end{equation}
is the sum of the Gaussian covariance \(\mathbf{Cov}^{\rm{G}}\) and the non-Gaussian covariance. The non-Gaussian covariance further decomposes into the connected non-Gaussian covariance \(\mathbf{Cov}^{\rm{cNG}}\) and the super-sample covariance \(\mathbf{Cov}^{\rm{SSC}}\). The exact formulation for these three covariance terms is detailed in Appendix A of \cite{cosmolike}. In this work, we drop the subdominant connected non-Gaussian covariance and keep the super-sample covariance \citep{Barreira2018_SSC, Takada2013_SSC}.

\subsection{Systematics Modeling\label{sec:sys}}
We model five types of systematic uncertainties: redshift uncertainty, multiplicative shear calibration bias, intrinsic alignment (IA), baryonic feedback, and galaxy bias. The related parameter and prior spaces are summarized in \tab{\ref{tab:prior}}. A detailed description of the systematics modeling is provided in \citep{cosmolike, Eifler_LSST_Roman}.

We follow \citep{Ma2006_photoz} to model the redshift uncertainty with a Gaussian model parameterized by the bias \(\Delta_z\) and the scatter \(\sigma_z\). We assume no bias in the fiducial cosmology, and take \(\sigma_z^\mrm{WL}=0.01\) and \(\sigma_z^\mrm{KL}=0.002\) \citep{Xu2023}.

We model the multiplicative bias as a Gaussian centered at 0 with standard deviation $\sigma_{m^i}$ per tomographic bin. We choose $\sigma_{m_i}=0.002$ for WL, supported by recent simulations \citep{Berlfein2026}. For KL, we account for the fact that KL galaxies generally have an image signal-to-noise ratio of 100 (see \fg{\ref{fig:KL_sample_char}}) and are spatially resolved. This implies a smaller multiplicative bias, which motivates our choice of $4\times 10^{-4}$. We note that shear calibration bias at either level is subdominant in the overall error budget.

We adopt the non-linear alignment model (NLA; see \citep{BridleKing2007, Hirata2004_IA, Krause2016_IA}) only for red elliptical galaxies used in WL, since KL is immune to IA and any IA-like systematics \citep{TED}. We model the IA amplitude \(A_\mrm{IA}\), the luminosity dependence \(\beta_\mrm{IA}\), and the redshift dependence \(\eta_\mrm{IA}\) and \(\eta_\mrm{IA}^{\mrm{high}\textit{-}z}\), where the fiducial values and choice of priors are listed in \tab{\ref{tab:prior}}. 

For galaxy-galaxy lensing and galaxy clustering, we refrain from modeling baryonic uncertainties because of conservative scale cuts and conservative marginalization over galaxy bias. We model the impact of baryonic physics on the angular shear-shear power spectrum using a principal component analysis (PCA) framework \citep{Eifler2015_baryon, HJHuang2019_baryon, HJhuang2021_baryon}. We derive the principal components \(\mathbf{PC}_i\) with a covariance-driven weighting factor using Horizon-AGN, TNG100, MassiveBlack-II, EAGLE, cosmo-OWLs, and BAHAMAS simulations \citep{Chisari2018_HorizonAGN, TNG100, MBH2, eagle, McCarthy2017_BAHAMAS, LeBrun2014_cOWL, vanDaalen2020_bahamas_cOWLS}. The cosmo-OWLs and BAHAMAS suites include three different AGN models that account for different AGN temperatures \citep{vanDaalen2020_bahamas_cOWLS}. The baryon-contaminated data vector is modeled by the linear combination of \(n\) principal components
\begin{equation}
    \mathbf{M}_\mrm{bary}(\bm{p}, \mathbf{Q})=\mathbf{M}_\mrm{dmo}(\bm{p})+\sum_{i=1}^{n} Q_i \mathbf{PC}_i,
\end{equation}
where \(\mathbf{Q}=(Q_i)_{i\le n}\) is the amplitude of \(\mathbf{PC}_i\) and $\mathbf{M}_\mrm{dmo}$ is the dark-matter-only data vector. We include the first three principal components in our analysis and use Gaussian priors centered at zero for \(Q_i\). For the first principal component, we set the prior so that the 68\% confidence interval matches the prediction from X-ray measurements \citep{Xu2025_6x2pt, Akino2022_xray}. This is motivated by the fact that the first principal component closely traces the gas mass fraction \citep{Xu2025_6x2pt}. For \(Q_2\) and \(Q_3\), we use a broad prior range that covers all hydrodynamic simulations within the 95\% confidence interval. The principal components depend on a galaxy sample's redshift distribution and the specific covariance matrix. We determine priors on the Q's for KL and WL using the same physical relation, yielding different values.    

We parameterize the galaxy bias as a linear function of the effective redshift \(b_1(z)=b_1^0z+b_1^1\). This differs from the more common approach of assuming one linear bias parameter per lens redshift bin. We choose this because this paper studies science cases with more tomographic bins, and we want to reduce the impact of additional systematic parameters that would otherwise affect our results. We adopt the linear relation estimated from fitting the High-z Emission Line Survey (HiZELS) \citep{Cochrane2017:bias, Merson2019_b1z} since star-forming galaxies dominate the KL sample. 

\section{Galaxy Sample Selection\label{sec:sample}}
The Medium Tier of the HLWAS will have $YJH$ imaging\footnote{$YJH$ corresponds to $F106$, $F129$, and $F158$. The COSMOS Mock Catalog has similar filters of $F110W$ and $F160W$.} and grism spectroscopy over 2,300 deg$^2$ with imaging $5\sigma$ point-source photometry depth of $Y=26.5$ and $J=H=26.4$ (AB) and $1.5\times 10^{-16}\,\mathrm{erg\,cm^{-2}\,s^{-1}}$ in grism ($5\sigma$ extended source line threshold, with half-light radius of $0.''3$) \citep{ROTAC2025}. This configuration is different from the one used in \citep{Xu2023}, which assumes a 2,000 deg\(^2\) survey with a \(5\sigma\) H-band depth of 26.54 (AB) and a grism sensitivity of \(10^{-16}\mrm{\, erg\,cm^{-2}\,s^{-1}}\) at a \(7\sigma\) point-source detection limit (see Fig. 2-15 in \citep{WFIRST_AFTA}), which corresponds to $(0.86-1.35)\times10^{-16}\mrm{\, erg\,cm^{-2}\,s^{-1}}$ for $5\sigma$ extended source of $r_{1/2}=0.^{\prime\prime}3$, depending on wavelength. We derive the KL and the WL samples for the new survey configuration below.

\subsection{Kinematic Lensing\label{sec:KL_sample}}
Following \citep{Xu2023}, we use the COSMOS Mock Catalog \citep{Jouvel2009} to obtain the galaxy population. To qualify for a KL measurement, a spiral galaxy needs high signal-to-noise emission-line measurements and a sufficiently large apparent size to measure the rotation curve. We therefore impose the following criteria: 
\begin{enumerate}
    \item At least one of the emission lines is resolved in the grism of resolution \(R_\lambda=451\frac{\lambda}{1\mu m}\) spanning from \(1-1.93\,\rm \mu m\).
    \item The resolved emission line flux is brighter than the $5\sigma$ extended source depth \(1.5\times 10^{-16}\mathrm{erg\,cm^{-2}\,s^{-1}}\)
    \item The apparent half-light radius is larger than Roman's point spread function, which is 0.1 arcsec.
    \item \(F110W\le26.5\) (AB)
\end{enumerate}

\citep{Xu2023} also applies the additional constraint of spatial resolution \(\mathcal{R}\ge 0.4\)\footnote{\citep{Xu2023} follows \citep{Hirata2012_ETC}'s definition on spatial resolution. They define \(\mathcal{R}\equiv \left(1 + EE50^2_\mrm{PSF}/r_\mrm{eff}^2\right)^{-1}\) \citep{HirataUros2003}, where \(EE50_\mrm{PSF}\) is the 50 percent effective radius of the point spread function and \(r_\mrm{eff}\) is the galaxy effective radius.} and a 5\(\sigma\) point source limit on F160W for sample selection. We perform the same procedure and find that these additional criteria do not alter the selection results. \fg{\ref{fig:KL_sample_char}} characterizes the distribution of the emission-line strength and the image signal-to-noise ratio of the KL sample in the Medium Tier.
Additionally, we further impose a 0.5 dropout rate to account for the observing inefficiency and completeness. These, in the end, result in a galaxy number density of 3.9~arcmin\(^{-2}\) for our likelihood analysis. For our later \(3\times2\)pt analysis, we assume the lens sample matches the source sample \citep{Fang2022_lenseqsrc, Boruah2024_lenseqsrc}.

\begin{figure}
    \centering
    \includegraphics[width=0.99\linewidth]{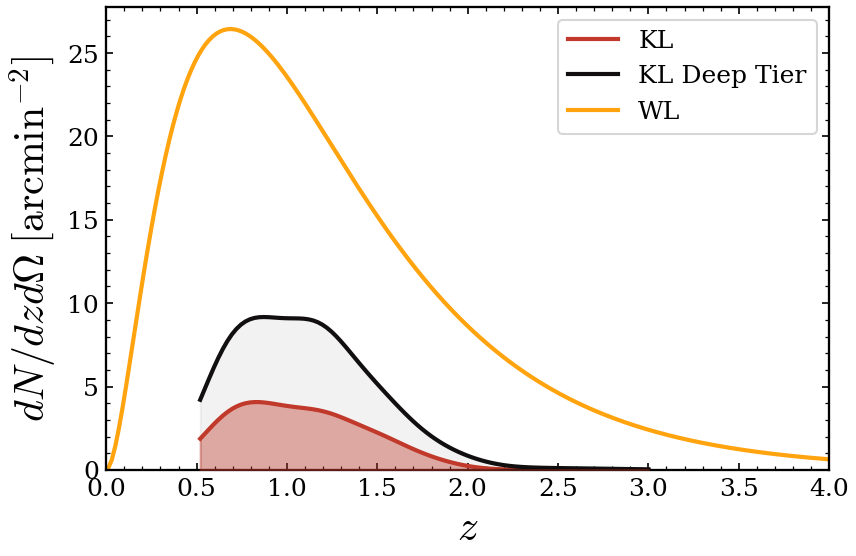}
    \caption{Redshift distribution of the galaxy samples from the updated configuration in the ROTAC report for WL (yellow), KL (red), and KL Deep Tier (black).\label{fig:dndz}}
\end{figure}

\begin{figure}
    \centering
    \includegraphics[width=\linewidth]{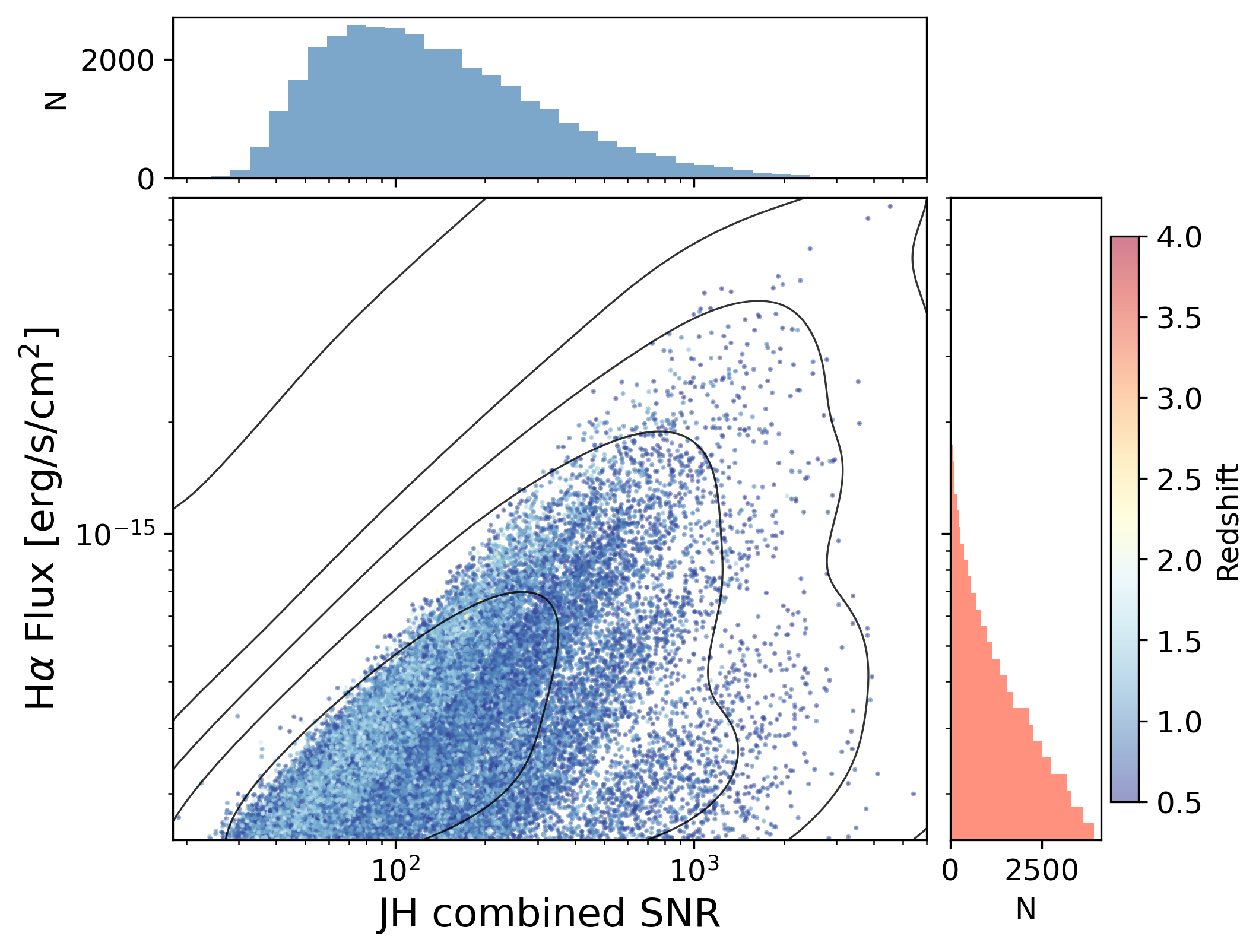}
    \caption{Characteristic $\mrm{H}\alpha$ emission flux and joint J+H band SNR of the Roman KL sample selected from the COSMOS Mock Catalog. We show selected galaxies as scatter points colored by redshift, and the smoothed kernel density estimate as black contours. The upper panel shows the 1D marginal histogram of J+H SNR, and the right panel shows $\mrm{H}\alpha$ flux.}
    \label{fig:KL_sample_char}
\end{figure}

We also estimate the redshift distribution for the Deep Tier following the recommendation in the ROTAC report and the HLWAS documentation\footnote{\url{https://roman-docs.stsci.edu}}, which has grism sensitivity of \(5.6\times 10^{-17} \mrm{erg\,cm^{-2}\,s^{-1}}\) and Y-band imaging depth of 27.7 (AB) in \(F110W\) over 19.2~deg\(^2\). The resulting sample redshift distribution is shown as the black curve in \fg{\ref{fig:dndz}}, and the galaxy number density is 9.65~arcmin\(^{-2}\).

\subsection{Weak Lensing\label{sec:WL_sample}}
We assume the Medium Tier as the reference WL survey and adopt the WL sample estimated for Data Challenge 1, initialized by the Roman Project Infrastructure Team (private communication). It follows a procedure similar to the previous Roman HLWAS forecasts (e.g.~\citep{Eifler2021_Roman_multiprobe}) by computing a realistic survey scenario with the Roman survey exposure time calculator \citep{Hirata2012_ETC} and then applying it to the CANDELS catalog \citep{Grogin2011_CANDELS} to obtain the sample. We fit the sample with the Smail redshift function \citep{Samil1994}, fixing \(\beta =2\). This estimation yields a redshift distribution in \fg{\ref{fig:dndz}} that peaks at \(z\sim0.7\). The total number density is 41.3~arcmin\(^{-2}\). We assume the WL shape noise $\sigma_\epsilon^\mrm{WL}=0.37$ per galaxy \citep{Chang2013}.

\section{Roll-angle Study\label{sec:roll}}

The roll angle that optimizes the trade-off between spectra decomposition and survey efficiency is between 8 and 17 degrees \citep{Nedkova}. \citep{Xu2023} estimates the shape noise of KL using a toy model, assuming a 90-degree roll-angle separation. 
Intuitively, since the line-of-sight velocity along the major and minor axes relates closely to the inclination and $g_\times$, one would expect that an orthogonal spectra configuration would optimize the cosmic shear measurement of KL, and a less-optimal spectra configuration would degrade the KL shape noise.
In this section, we examine how the roll-angle separation of grism exposures affects KL shear uncertainty through mock observations. The metric we focus on is the correlation coefficient between the model parameter pairs, $g_+$ v.s. inclination $\cos i$, and $g_\times$ v.s. intrinsic position angle $\theta_\mathrm{int}$, in the Fisher information matrix calculated from a suite of mock image and grism observations. The parameter pairs are chosen to be fully degenerate with each other when only images are considered. We describe the mock observation modeling in Sect.~\ref{sec:kltools} and show Fisher information results in Sect.~\ref{sec:roll_angle_fisher}. We show two particular Fisher information examples in Appendix~\ref{sec:append_FIM} to build intuition for how grism data helps break the degeneracy between shear and shape at different roll-angle separations. 

\begin{figure*}
    \centering
    \includegraphics[width=\linewidth]{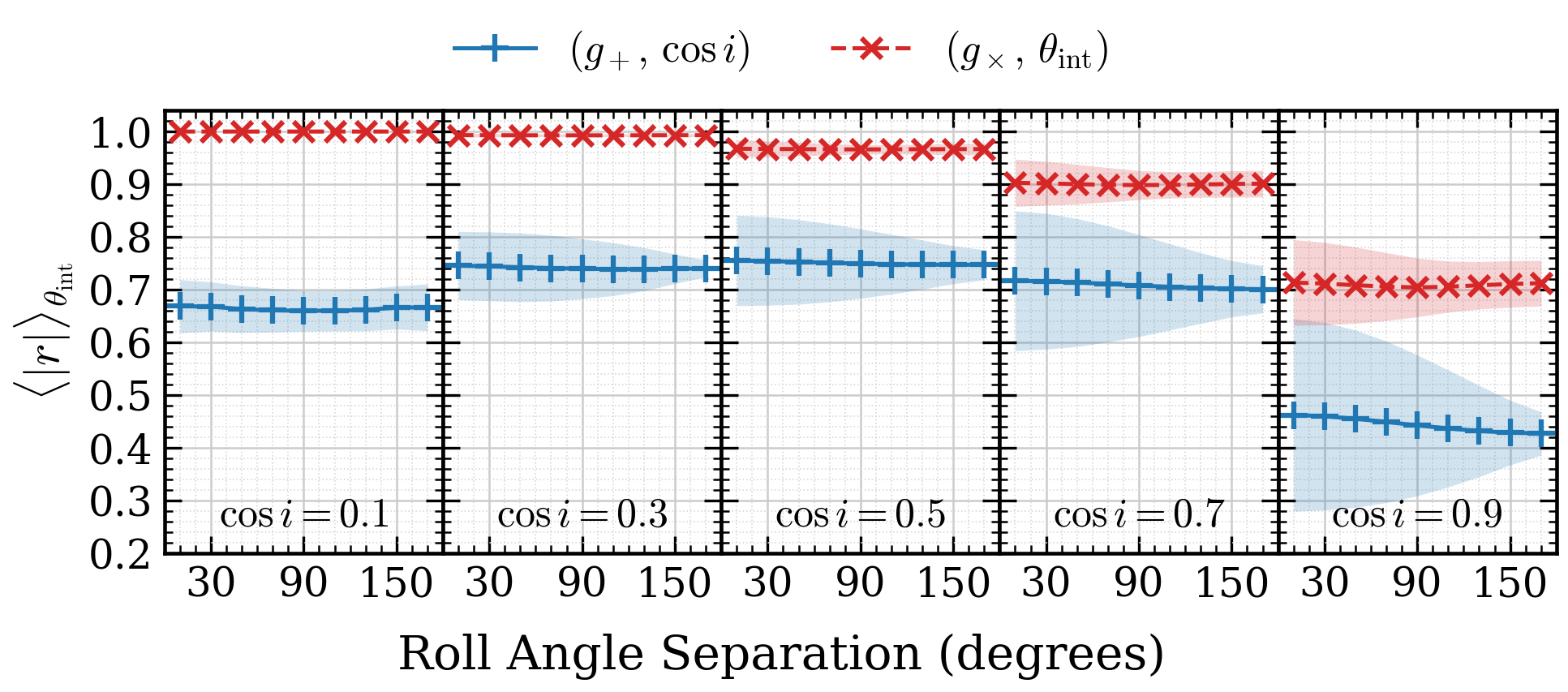}
    \caption{The correlation coefficients between degenerate parameter pairs, $(g_+,\,\cos i)$ (blue) and $(g_\times,\,\theta_\mathrm{int})$ (red), as a function of roll angle separation between the two grism observations. We break down the roll angle dependency into five inclination bins from left ($\cos i=0.1$, close to edge-on) to right ($\cos i=0.9$, close to face-on). The correlation coefficients in each panel are averaged over different position angles in $[0\deg,\,360\deg)$. The shaded regions show the standard deviation of the roll angle dependency for various position angles.}
    \label{fig:corr_coeff_roll_angle}
\end{figure*}

\subsection{Mock KL Measurement\label{sec:kltools}}
We use \textsc{kltools}\footnote{\url{https://github.com/KinematicLensing/kl-tools}} to build mock Roman image and grism data. We refer to~\citep{Xu2024} for a detailed description of \textsc{kltools} for equivalent fiber-spectra-based KL measurement, and summarize the difference between a grism-based measurement below.

We consider three independent cutouts for each mock galaxy: one broadband image from the imaging survey, and two grism images cut around $\mathrm{H}\alpha$ emission, observed by the grism survey. 
The roll angles of the two grism cutouts differ by $\Delta\theta_\mathrm{RA}$, and we align the dispersion direction of the first grism image with the $x$-axis of the cutout. 
We assume independent Gaussian pixel noise for both the broadband image and grism cutouts. 

\textsc{kltools} first generates a dimensionless intensity profile $I(\bm{x}|\bm{\theta})$ and a line-of-sight (LoS) rotation velocity field $v_\mathrm{LoS}(\bm{x}|\bm{\theta})$ ($\mathrm{km}\,\mathrm{s}
^{-1}$) given the model parameter vector $\bm{\theta}$, and the cutout coordinate grid $\bm{x}$. We note that both the intensity profile and the velocity field include shear distortions. We adopt an inclined exponential disk profile for $I(\bm{x}|\bm{\theta})$, and the 3D rotation velocity is modeled as \citep{Courteau1997_arctan}
\begin{equation}
    V(r)=\frac{V_\mathrm{circ}}{\pi/2}\tan^{-1}(r/R_\mathrm{vscale}),
\end{equation}
where $V_\mathrm{circ}$ is the maximum rotation velocity and $R_\mathrm{vscale}$ is the scale radius of the rotation curve.

The observer-frame spectral energy distribution (SED) $F_\lambda^\mathrm{obs}(\lambda|\bm{\theta})$ ($\mathrm{erg\,s^{-1}\,cm^{-2}\,nm^{-1}}$) is also calculated given $\bm{\theta}$ and observer-frame wavelength $\lambda$. We model the SED around the $\mathrm{H}\alpha$ emission line as a Gaussian line profile with zero continuum, assuming perfect continuum subtraction.

The 3D photon distribution is constructed as 
\begin{equation}
    f(\bm{x},\lambda|\bm{\theta})=I(\bm{x}|\bm{\theta})F_\lambda^\mathrm{obs}\left(\left.\frac{\lambda}{1+v_\mathrm{LoS}(\bm{x})/c}\right|\bm{\theta}\right)\frac{\lambda}{hc}.
\end{equation}
The broadband image $\bm{X}_\mathrm{I}(\bm{x}|\bm{\theta})$ can be expressed as 
\begin{equation}
    \bm{X}_\mathrm{I}(\bm{x}|\bm{\theta})=\frac{At}{G}(\mathcal{P}*I)(\bm{x})\int\mathrm{d}\lambda\frac{F_\lambda^\mathrm{obs}(\lambda|\bm{\theta})}{hc/\lambda}\mathcal{T}(\lambda),
\end{equation}
where $\mathcal{P}$ is the point-spread function of Roman, $\mathcal{T}$ is the total system throughput, $A$ is the telescope aperture area, $t$ is the exposure time, $G$ is the detector gain, $*$ is convolution, and $h$ is the Planck constant.

The grism image $\bm{X}_\mathrm{G}(\bm{x}|\bm{\theta})$ can be expressed as 
\begin{equation}
    \bm{X}_\mathrm{G}(\bm{x}|\bm{\theta})=\frac{At}{G}\int\mathrm{d}\lambda[(\mathcal{P}*f)(\bm{x}-\bm{d}(\lambda),\lambda|\bm{\theta})]\mathcal{T}(\lambda),
\end{equation}
where $\bm{d}(\lambda)$ is the dispersion law of the Roman grism. $\bm{d}(\lambda)$ is generally expressed as 
\begin{equation}
    \bm{d}(\lambda)=(x_0+dx(x_0,y_0,\lambda),\,y_0+dy(x_0,y_0,\lambda))^\mathrm{T},
\end{equation}
where $(x_0,\,y_0)$ is the position of the galaxy in the direct image, and $dx$ and $dy$ are polynomials calibrated from commissioning or laboratory data. Practically, because the grism cutouts span only a small wavelength range, we keep only the linear term in the polynomial trace around emission lines. 

We note that although we only consider $g_+$, $g_\times$, $\cos i$, $\theta_\mathrm{int}$, and the image and $\mathrm{H}\alpha$ half-light radius (hlr) $R_{h,\,\mathrm{bb}}$, $R_{h,\,\mathrm{H}\alpha}$ in Sect.~\ref{sec:roll_angle_fisher}, \textsc{kltools} supports a much larger parameter space of the model parameter vector $\bm{\theta}$, producing realistic model broadband images and 2D grism spectra. We fix other model parameters to their fiducial values, including $V_\mathrm{circ}=251.5\,\mathrm{km\,s}^{-1}$, $R_\mathrm{vscale}=0.^{\prime\prime}05$, intrinsic $\mathrm{H}\alpha$ width $\sigma_{\mathrm{H}\alpha}=0.175\,\mathrm{nm}$; image and grism spectra fluxes are fixed to an image SNR of 100 and grism spectra SNR of 50 each.

\subsection{Fisher Information and Shear Uncertainty Dependency\label{sec:roll_angle_fisher}}
Assuming independent Gaussian likelihoods for each observation, the Fisher information matrix can be written as 
\begin{equation}
    \mathcal{F}_{ij}\equiv \sum_{n=1}^{N_\mathrm{obs}} \frac{\partial \bm{X}_n}{\partial \theta_i}\bm{C}^{-1}_n\frac{\partial \bm{X}_n}{\partial\theta_j},
\end{equation}
where $\bm{X}_n$ is the n-th observation in the image and grism cutouts, and $\bm{C}_n$ is the corresponding pixel noise covariance matrix, which is diagonal assuming uncorrelated pixel noise. The conditional correlation coefficient $r(\theta_i,\,\theta_j)$ between two model parameters $\theta_i$ and $\theta_j$ is
\begin{equation}
    r(\theta_i,\,\theta_j)\equiv \frac{\mathcal{F}_{ij}}{\sqrt{\mathcal{F}_{ii}\,\mathcal{F}_{jj}}}.
\end{equation}
Note that when only broadband image is available, the $g_+$-$\cos i$ and $g_\times$-$\theta_\mathrm{int}$ are perfectly degenerate, $r(g_+,\,\cos i)=-1$, $r(g_\times,\,\theta_\mathrm{int})=1$. 
Adding a 2D grism spectrum breaks this degeneracy to $|r|<1$. Adding a second 2D grism spectrum with a different dispersion direction further reduces the correlation, and we investigate the dependence of $|r|$ on the roll-angle separation between the two grism exposures in Fig.~\ref{fig:corr_coeff_roll_angle}, leaving more detailed intuition in Appendix~\ref{sec:append_FIM}. Galaxies with different intrinsic position angles experience different roll angle dependencies; therefore, we consider the intrinsic position angle-averaged correlation coefficient $\langle|r|\rangle_{\theta_\mathrm{int}}$ (lines with markers) in Fig.~\ref{fig:corr_coeff_roll_angle}, with the shaded region showing the spread among different position angles. The face-on scenarios have the lowest correlation coefficient between degenerate parameter pairs, consistent with \citep{RS2023_shape_noise, Xu2024}. The $g_+$ and $g_\times$ components have different characteristic behavior. At high inclination (low $\cos i$), $g_+$ has flat dependency on $\Delta\theta_\mathrm{RA}$, and $\Delta\theta_\mathrm{RA}=90\deg$ is slightly preferred. However, at low inclination, the $\Delta\theta_\mathrm{RA}$ dependence is more obvious and decreases monotonically toward larger $\Delta\theta_\mathrm{RA}$. $g_\times$, by contrast, shows a consistently flat dependence across inclinations and prefers $\Delta\theta_\mathrm{RA}=90\deg$. When weighted by inverse variance in the real two-point correlation function measurement, the high $\cos i$ scenarios have much higher weight than the edge-on cases. Another factor is the spread of the roll-angle dependency. For $\Delta\theta_\mathrm{RA}=180\deg$, both $g_+$ and $g_\times$ have tight spread in $|r|$-$\Delta\theta_\mathrm{RA}$ relation. This means a more homogeneous and healthy weight for galaxies across different position angles. Therefore, $\Delta\theta_\mathrm{RA}=180\deg$ is preferred, given both the inverse-variance-weighted effective trend and the homogeneity of shear uncertainty across position angles.

The reference exposure time of the Medium Tier grism is 4 passes $\times$ 2 dithers per pass $\times$ $190\,\mathrm{s}$ per exposure, and the four passes have K-sequence roll angles around (0, 10, 170, 180) degrees~\citep{wang_hlss_2022}. This roll-angle setup also optimizes KL shear uncertainty, as discussed, and we conclude that KL does not require changes to the current reference roll-angle setting.

From our roll angle study simulations, we estimate an overall shape noise level of $\sigma_\epsilon^\mrm{KL}=0.05$ per galaxy for this galaxy sample. This corresponds to a rotational-velocity measurement uncertainty of $\sigma_v \sim 25-30 \, \rm{km \, s^{-1}}$, which is more conservative than assumed in \cite{Xu2023}. Studies with the Hubble Space Telescope have already achieved a precision of $\sigma_v = 15 \, \rm{km \, s^{-1}}$ using multiple roll angles \citep{OC20}.

\section{Forecasting the new survey strategy\label{sec:forecast}}

Given our galaxy sample selection in Sect. \ref{sec:sample} and the simulations that motivate the KL shape noise level for the roll-angle strategy of the Medium Tier (Sect. \ref{sec:roll}), we compute our data vector and covariance matrix as explained in Sect. \ref{sec:theory}.

\begin{figure*}
    \centering
    \includegraphics[width=0.99\linewidth]{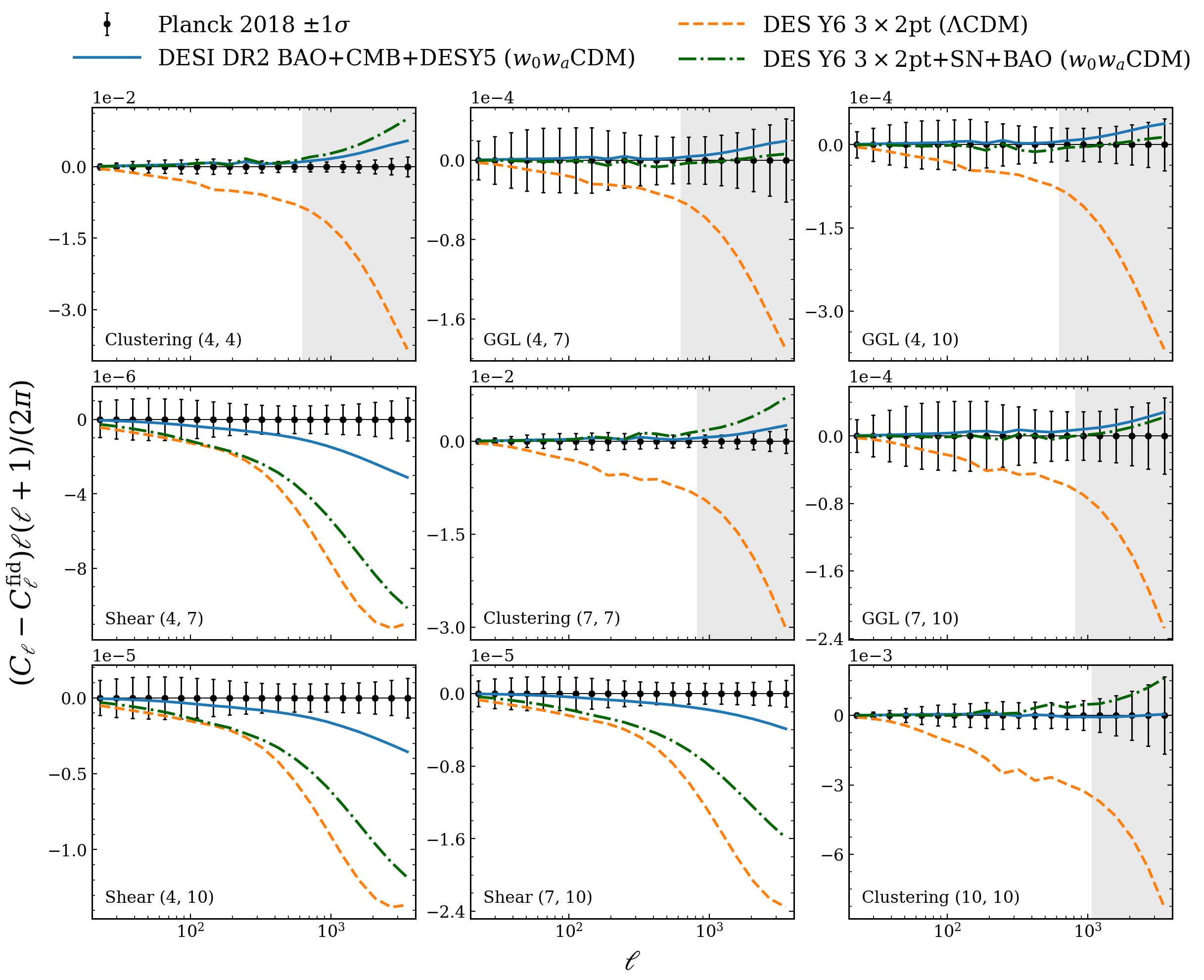}
    \caption{We compare the KL-$3\times2$pt data vector difference among different cosmological models from DESI DR2 BAO+CMB+DESY5 in \wcdm{} (blue solid curves), DES Y6 $3\times2$pt in \lcdm{} (orange dashed curves), and DES Y6 $3\times2$pt+SN+BAO in \wcdm{} (green dotted-dashed curves). The fiducial cosmology for this comparison is the Planck 2018 \lcdm{} prediction (black), and the error bar is estimated from the diagonal terms of the covariance matrix. The cosmic shear data vector occupies the lower triangle, galaxy-galaxy lensing the upper triangle, and the clustering the diagonal panels. Panel annotations denote the probe and the tomographic bin indices. The gray shaded regions are the scales excluded by \(R_{\min}=21 \,\mrm{Mpc}/h\).\label{fig:model-diff}}
\end{figure*}

\fg{\ref{fig:model-diff}} shows the difference of the simulated input KL-$3\times2$pt data vector (black dots) computed at Planck 2018 cosmology \citep{Planck2018} and best-fit cosmologies from recent experiments: 
\begin{itemize}
\item DESI DR2 BAO+CMB+DESY5 in \wcdm{} \citep{DESI_DR2_BAO} (blue)
\item DES Y6 $3\times2$pt in \lcdm{} \citep{DES2026:Y6_3x2} (orange)
\item DES Y6 $3\times2$pt+SN+BAO in \wcdm{} \citep{DES2026:DE} (green)
\end{itemize}
Error bars are obtained as the square root of the diagonal elements of the KL covariance matrix, and the shaded region indicates scale cuts in galaxy clustering and galaxy-galaxy lensing. We see strong differences in the models on small scales in the cosmic shear part of the data vector. These differences significantly exceed the error bars, indicating that we can best discriminate between models on these scales if we can control systematics, particularly baryonic physics. 

\begin{figure}
    \centering
    \includegraphics[width=0.99\linewidth]{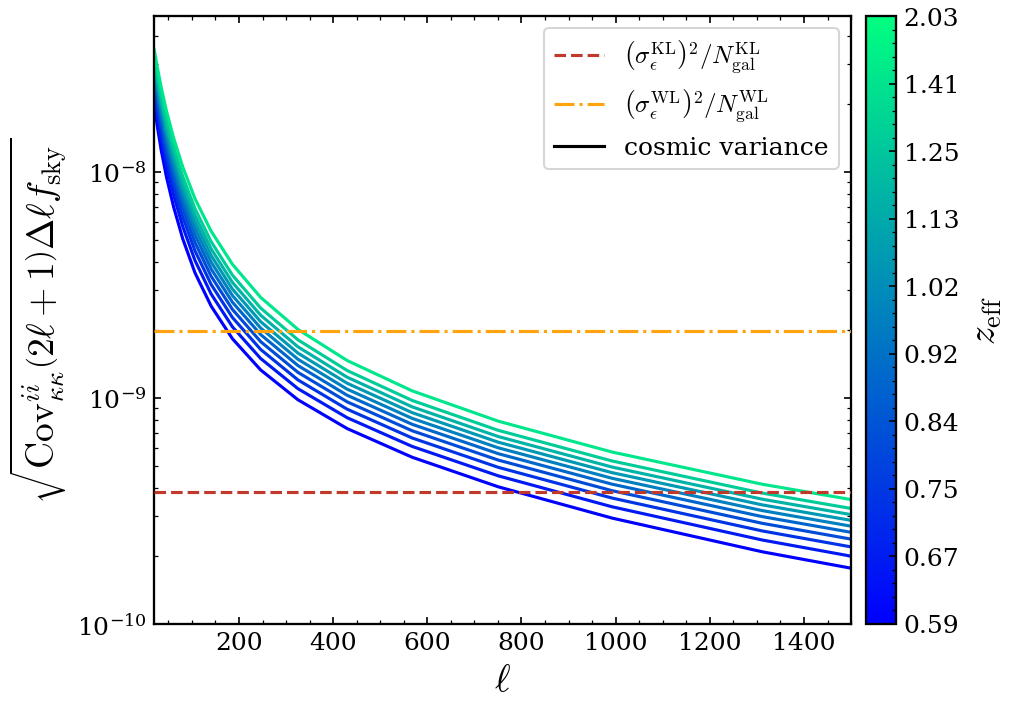}
    \caption{Comparison of cosmic variance at different effective redshifts (solid curves) and shape noise for KL (red dashed line) and WL (yellow dashed line). The $y$-axis is the diagonal part of the covariance matrix scaled to keep the shape noise constant over $\ell$, and the $x$-axis is the multipole \(\ell\).  KL pushes the cosmic variance-dominated regime toward smaller scales compared to WL at the same effective redshift.\label{fig:noise_contribution}}
\end{figure}

We further explore the difference between WL and KL in the shape-noise contribution over the total uncertainty budget in \fg{\ref{fig:noise_contribution}}. Both scenarios assume the fiducial forecast setup with 10 tomographic bins over the Medium Tier. At the same cosmic variance, KL is less dominated by shape noise than WL. While this opens the exciting opportunity to constrain cosmological parameters and different cosmological models more tightly, it also requires better modeling of baryonic physics and other small-scale cosmic shear uncertainties. 

We focus on two aspects of the KL forecasts: (i) updating the KL forecast of \citep{Xu2023} with the ROTAC Medium Tier survey configuration (Sect.~\ref{sec:KL-vs-WL}), and (ii) comparing the KL constraining power under different survey tiers and data timelines, including the performance of KL with the Y1 data (Sect.~\ref{sec:tiers}). For each forecast, we run a standard \lcdm{} and a \wcdm{} scenario. Our choice of priors and the fiducial values are summarized in \tab{\ref{tab:prior}}. Our baseline forecast setting adopts 10 tomographic bins and 20 multipole bins logarithmically spaced between \(\ell_\mrm{\min}=20\) and \(\ell_\mrm{\max}=4{,}000\). 

\begin{table}
    \centering
    \caption{Summary of the survey design used in this work. The galaxy number density was estimated in Sect.~\ref{sec:KL_sample}\label{tab:area_ngal}}
    \setlength{\tabcolsep}{8pt}
    \begin{tabular}{ccc}
    \toprule
    Design  & Area (deg\(^2\)) & \(N_\mrm{gal}\) (arcmin\(^{-2}\)) \\
    \midrule
    WL      & 2300  & 41.3  \\
    KL      & 2300  & 3.9 \\
    KL Deep Tier & 19.2  & 9.65  \\
    KL Y1   & 700   & 3.9 \\
    \toprule
    \end{tabular}
\end{table}

\begin{table*}
	\centering
    \caption{Summary of fiducial parameters and the priors for different scenarios in \CL{}simulations. We only perform \(3\times2\)pt analyses with KL, and hence we do not list any galaxy bias parameterization for WL\label{tab:prior}.}
	\begin{tabular}{ccccc}
	\toprule
	Cosmology Parameters & \multicolumn{2}{c}{Fiducial} & \multicolumn{2}{c}{Prior}\\
	\midrule
	$\Omega_\mrm{m}$&\multicolumn{2}{c}{0.3156}&\multicolumn{2}{c}{$\Udist{0.095}{0.585}$}\\
	$\sigma_8$&\multicolumn{2}{c}{0.831}&\multicolumn{2}{c}{$\Udist{0.5}{1.1}$}\\
	$n_s$&\multicolumn{2}{c}{0.9645}&\multicolumn{2}{c}{$\Udist{0.84}{1.06}$}\\
	$\Omega_\mrm{b}$&\multicolumn{2}{c}{0.0491685}&\multicolumn{2}{c}{$\Udist{0.005}{0.095}$}\\
	$h_0$&\multicolumn{2}{c}{0.6727}&\multicolumn{2}{c}{$\Udist{0.4}{0.9}$}\\
	$w_0$&\multicolumn{2}{c}{-1}&\multicolumn{2}{c}{$\Udist{-2.1}{0}$}\\
	$w_a$&\multicolumn{2}{c}{0}&\multicolumn{2}{c}{$\Udist{-2.6}{2.6}$}\\
	\toprule
    
	\multirow{2}{*}{Systematic Parameters} & \multicolumn{2}{c}{WL} & \multicolumn{2}{c}{KL}\\
	 & Fiducial & Prior & Fiducial & Prior \\ 
	\midrule
	$\Delta_{z}^{i}$&0.0&$\Ndist{0}{2}{-3}$&0.0&$\Ndist{0}{4}{-4}$\\
	$\sigma_z^{i}$&0.01&$\Ndist{0.01}{2}{-3}$&0.002&$\Ndist{0.002}{4}{-4}$\\
	$m^{i}$&0.0&$\Ndist{0}{2}{-3}$&0.0&$\Ndist{0}{4}{-4}$\\
    $b_1^0$ &-&-& 0.98 & $\mathcal{U}(0.0,\, 1.7)$ \\
    $b_1^1$ &-&-& 0.34 & $\mathcal{U}(0.1,\, 1.0)$ \\
	$A_{\mrm{IA}}$&5.92&$\mathcal{N}(5.92,\,3.0)$&-&-\\
	$\beta_{\mrm{IA}}$&1.1&$\mathcal{N}(1.1,\,1.2)$&-&-\\
	$\eta_{\mrm{IA}}$&-0.47&$\mathcal{N}(-0.47,\,3.8)$&-&-\\
	$\eta_{\mrm{IA}}^{\mrm{high}\textit{-}z}$&0.0&$\mathcal{N}(0.0,\,2.0)$&-&-\\
	$Q_1$&0.0&$\mathcal{N}(0.0,\,5.3)$&0.0&$\mathcal{N}(0.0,\,16.78)$\\
	$Q_2$&0.0&$\mathcal{N}(0.0,\,3.24)$&0.0&$\mathcal{N}(0.0,\,5.29)$\\
	$Q_3$&0.0&$\mathcal{N}(0.0,\,0.97)$&0.0&$\mathcal{N}(0.0,\,1.59)$\\
	\bottomrule
	\end{tabular}
\end{table*}

\begin{figure*}
\centering
    \includegraphics[width=0.49\linewidth]{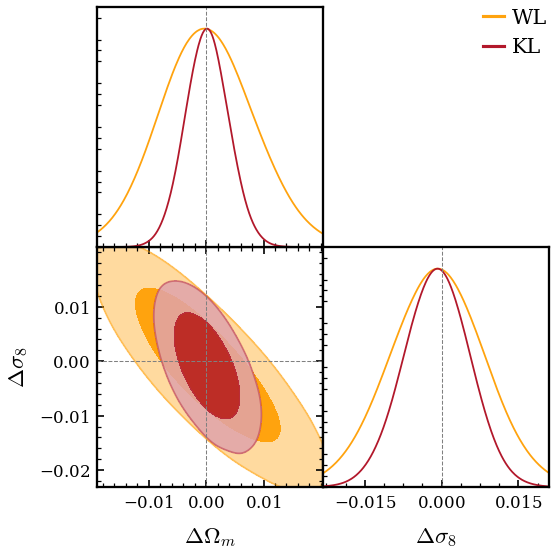}
    \includegraphics[width=0.49\linewidth]{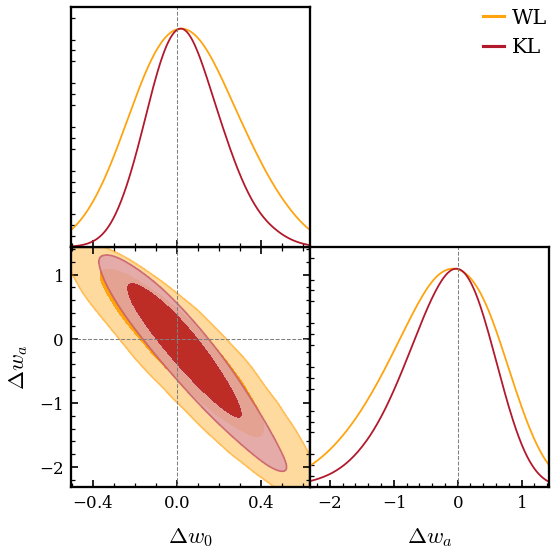}
    \caption{\textit{Left:} Comparison of the WL Medium Tier (yellow) and the KL Medium Tier design (red) for \(\Omega_{\rm m}\text{--}\sigma_8\) in the \lcdm{} parameter space. This figure uses 68\% and 95\% contour levels, as in the rest of the paper unless otherwise stated. \textit{Right:} the same comparison as the left panel but for \(w_0\text{--}w_a\) in the \wcdm{} parameter space. \label{fig:old_new_KL}}
\end{figure*}

\subsection{Weak Lensing and Kinematic Lensing in \texorpdfstring{$\Lambda$CDM}{PDFstring} and \texorpdfstring{$w_0w_a$CDM}{PDFstring}\label{sec:KL-vs-WL}}
As shown in \tab{\ref{tab:area_ngal}}, the new survey configuration returns a total number density of the KL sample of 3.9~arcmin\(^{-2}\) over a survey area of 2,300~deg\(^2\). The WL galaxy sample in the new survey design comes out to 41.3 ~arcmin\(^{-2}\), also over the same area. Compared to the previous study \cite{Xu2023}, we see a reduced number density because the new survey design uses a shallower, wider survey strategy. 

\begin{table*}
    \centering
    \caption{Summary of marginalized parameter constraints and the figure of merit in the \lcdm\label{tab:LCDM_1D}}
    \setlength{\tabcolsep}{8pt}
    \begin{tabular}{ccccccc}
        \toprule
        Dataset & $\Omega_\mrm{m}$ & $\sigma_8$ & $n_s$ & $\Omega_\mrm{b}$ & $h_0$ & $\mrm{FoM}_{\Omega_\mrm{m}\sigma_8}$\\
        \midrule
        WL              & 0.00829   & 0.00929   & 0.02826   & 0.01965   & 0.09675   & 22200.46    \\
        KL              & 0.00378   & 0.00642   & 0.02689   & 0.01902   & 0.09302   & 49856.78    \\
        KL Deep Tier    & 0.02376   & 0.03093   & 0.06179   & 0.02634   & 0.13567   & 3058.13     \\
        KL Y1           & 0.00643   & 0.00966   & 0.03187   & 0.02340   & 0.11181   & 24412.32    \\
        KL-$2\times2$pt & 0.01134   & 0.01511   & 0.04047   & 0.00462   & 0.04740   & 10188.61    \\
        KL-$3\times2$pt & 0.00304   & 0.00539   & 0.02233   & 0.00488   & 0.03443   & 71017.35    \\
        \bottomrule
    \end{tabular}
\end{table*}

Similar to the old survey configuration, the new configuration shows a significant improvement in KL over WL (see \fg{\ref{fig:old_new_KL}}). In \lcdm, KL shrinks the 1D marginalized error bar for \(\Omega_\mrm{m}\) and \(\sigma_8\) by a factor of 2.2  and 1.4, respectively, and it improves \(\mrm{FoM}_{\Omega_\mrm{m}\sigma_8}\)\footnote{We define the figure of merit of parameters \(p_1\) and \(p_2\) as \(\mrm{FoM}_{p_1,p_2}=\left[\det \mrm{Cov}(p_1,\,p_2)\right]^{-1/2}\), where the parameter covariance \(\mrm{Cov}(p_1,\,p_2)\) is estimated from the posterior of \(p_1\) and \(p_2\).} by a factor of 2.2 (\tab{\ref{tab:LCDM_1D}}). In \wcdm, we find a 2.1 times improvement in constraining power in \(\mrm{FoM}_{w_0w_a}\) (see \tab{\ref{tab:wCDM_1D}}). Overall, we find a slightly less significant improvement of KL over WL compared to the old survey design \citep{Xu2023}. This is also partly because this paper treats baryonic physics more realistically, which affects KL more than WL.

\begin{table*}
    \centering
    \caption{Summary of marginalized parameter constraints and the dark energy figure of merit in the \(w_0 w_a\)CDM\label{tab:wCDM_1D}}
    \setlength{\tabcolsep}{8pt}
    \begin{tabular}{ccccccccc}
        \toprule
        Dataset & $\Omega_\mrm{m}$ & $\sigma_8$ & $n_s$ & $w_0$	& $w_a$ &	$\Omega_\mrm{b}$ &	$h_0$ & $\mrm{FoM}_{w_0w_a}$\\
        \midrule
        WL              & 0.02546   & 0.02534   & 0.03160   & 0.25327   & 0.85267   & 0.02042   & 0.09752   & 10.9526   \\
        KL              & 0.01514   & 0.01859   & 0.03029   & 0.17657   & 0.67906   & 0.01987   & 0.09586   & 22.5411   \\
        KL Deep Tier    & 0.04499   & 0.04808   & 0.06667   & 0.47433   & 1.47330   & 0.02594   & 0.13793   & 2.4493    \\
        KL Y1           & 0.02252   & 0.02642   & 0.03511   & 0.26719   & 1.00195   & 0.02416   & 0.11168   & 10.0753   \\
        KL-$2\times2$pt & 0.03812   & 0.03315   & 0.04522   & 0.35348   & 1.07038   & 0.00750   & 0.06130   & 7.5095    \\
        KL-$3\times2$pt & 0.01289   & 0.01501   & 0.02372   & 0.15566   & 0.54830   & 0.00541   & 0.03972   & 34.5773   \\
        \bottomrule
    \end{tabular}
\end{table*}

\begin{figure}
    \centering
    \includegraphics[width=0.99\linewidth]{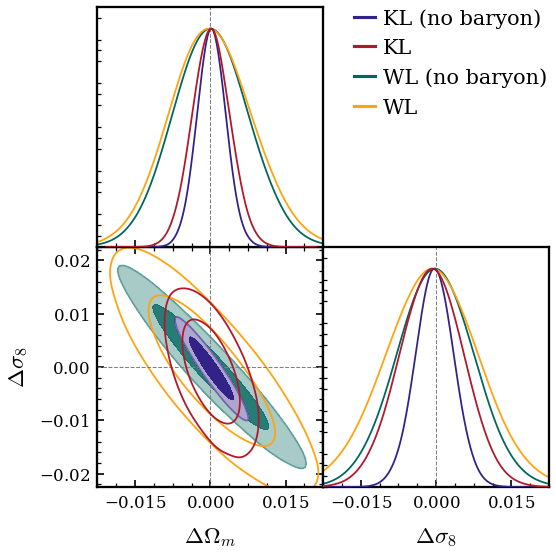}
    \caption{The impact of baryonic feedback on KL and WL in \lcdm{} cosmology. The red and yellow contours are KL and WL with baryonic systematics. The dark blue and green contours are KL and WL with perfect knowledge of baryonic physics. \label{fig:baryon-forecast}}
\end{figure}

We further explore the importance of baryonic physics in \fg{\ref{fig:baryon-forecast}}. KL with perfectly calibrated baryonic systematics gives 5.3 times better constraints in \(\mrm{FoM}_{\Omega_\mrm{m}\sigma_8}\) and 2.5 times better constraints in \(\mrm{FoM}_{w_0w_a}\) compared to KL with baryonic systematics, as they can be controlled today. In comparison, baryonic systematics degrade WL's constraining power by 2.5 times in  \(\mrm{FoM}_{\Omega_\mrm{m}\sigma_8}\) and 1.3 times in \(\mrm{FoM}_{w_0w_a}\). As expected, KL is more sensitive to baryonic feedback than WL because of smaller shape-noise uncertainties on small scales (\fg{\ref{fig:noise_contribution}}).  This opens up a promising opportunity to improve KL constraining power through better modeling and cross-calibration of baryonic physics. 

\subsection{KL performance for different survey tiers\label{sec:tiers}}
We forecast a 700~deg\(^2\) survey at Medium Tier depth, which Roman will have observed approximately by the end of 2028\footnote{\href{https://roman-docs.stsci.edu/roman-community-defined-surveys/roman-observations-in-the-first-two-years-of-science-operations}{Roman Observations in the First Two Years of Science Operations}}. We refer to this dataset as Y1 data. In this section, we forecast the KL performance on Y1 data (KL Y1) and compare with the full Medium Tier. Throughout the discussion, we refer to the full dataset as the Y5 data, since the Roman primary mission is scheduled for five years, although we note that the Medium Tier field will likely have been observed before the five-year timeline. 

In addition to the Medium Tier data, the Deep Tier, which will have completed observations by early 2028, will be highly useful for KL. We have estimated the KL sample for the Deep Tier in Sect.~\ref{sec:KL_sample} as the black redshift distribution in \fg{\ref{fig:dndz}}. For the Y1 data, we assume it shares the same detection sensitivity as the Y5 data. That is, it shares the same redshift distribution and galaxy number density as our estimation in Sect.~\ref{sec:KL_sample}, but covers only 700~deg$^2$ as noted in \tab{\ref{tab:area_ngal}}. We adopt the same tomographic bins, multipole bins, and the fiducial values as the Y5 KL forecast. 

\begin{figure}
    \centering
    \includegraphics[width=0.99\linewidth]{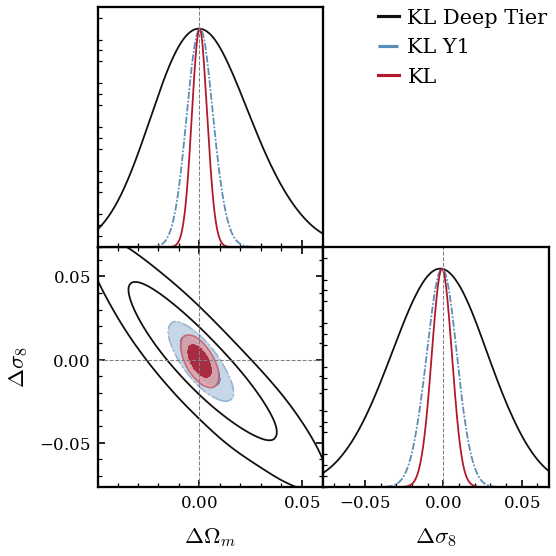}
    \caption{The $\Omega_\mrm{m}\text{--}\sigma_8$ KL forecasts for Roman Deep Tier (black), Y1 Medium Tier (blue), and Y5 Medium Tier (red). The forecast is in \(\Lambda \)CDM.\label{fig:tiers}}
\end{figure}

We compare the KL performance of Deep, Y1, and Y5 and summarize the results in \tab{\ref{tab:LCDM_1D}} (also see \fg{\ref{fig:tiers}}) and \tab{\ref{tab:wCDM_1D}}. The 1D marginalized error bar of \(\Omega_\mrm{m}\) (\(\sigma_8\)) is 0.00378 (0.00642), 0.00643 (0.00966), and 0.02376 (0.03093) for Y5, Y1, and the Deep Tier, respectively, and the corresponding \(\mrm{FoM}_{\Omega_\mrm{m}\sigma_8}\) is 49,857, 24,412, and 3,058. In \wcdm{}, \(\mrm{FoM}_{w_0w_a}\) is 22.54, 10.08, and 2.45 for Y5, Y1, and the Deep Tier. 
Even though the Deep Tier has a much higher galaxy number density, its \(\mrm{FoM}_{\Omega_\mrm{m}\sigma_8}\) is 16.3 times lower than the Y5 constraint, and its $w_0\text{--}w_a$ is prior-dominated. The \(\mrm{FoM}_{\Omega_\mrm{m}\sigma_8}\) of Y1 is 1.1 times larger than the WL with full Medium Tier data and has comparable \(\mrm{FoM}_{w_0w_a}\), while Y1 only has 700~deg\(^2\) compared to 2,300~deg\(^2\) of the Y5 data, demonstrating the reduction in shape noise.

\section{\texorpdfstring{$\mathbf{KL}\text{-}3\times2\mrm{pt}$}{PDFstring} analysis strategies \label{sec:multiprobe}}
In this section, we explore replacing WL with KL in the $3\times2$pt framework. We follow the same fiducial analysis setup as in Sect.~\ref{sec:forecast} with 10 tomographic bins and 20 multipole bins. In Sect.~\ref{sec:NL}, we quantify the information gain as we push toward narrower tomographic bins. We also quantify the impact of the Limber approximation in a narrow tomographic scenario. 

\begin{figure*}[htb]
\centering
    \includegraphics[width=0.49\linewidth]{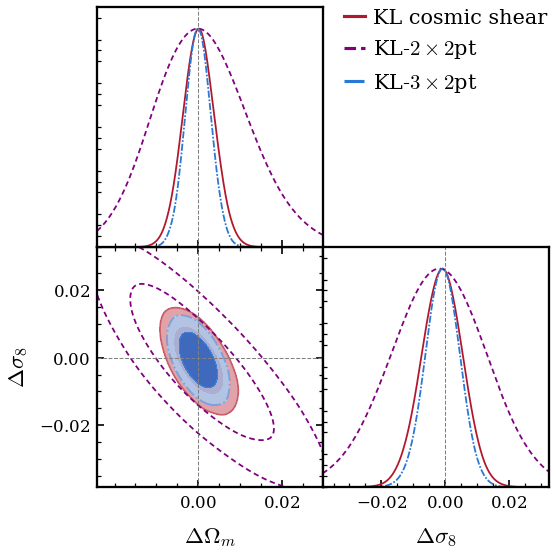}
    \includegraphics[width=0.49\linewidth]{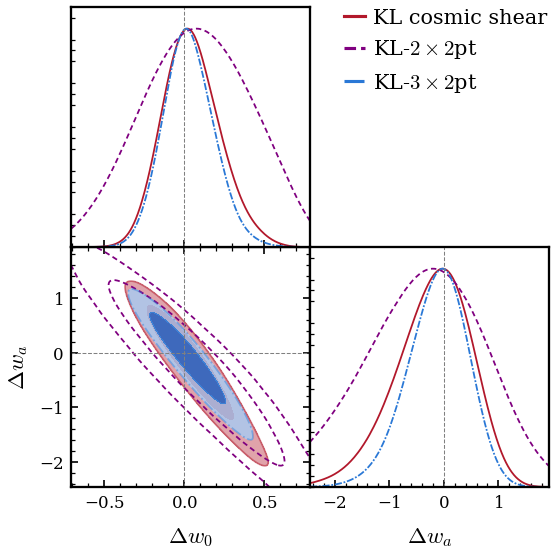}
    \caption{\textit{Left:} Comparison among KL cosmic shear (red), KL-$2\times2$pt (purple), and KL-$3\times2$pt (blue) for $\Omega_{\rm m}\text{--}\sigma_8$ in \lcdm{}. \textit{Right:} The same comparison as the left panel but for $w_0\text{--}w_a$ in \wcdm. KL-\(2\times2\)pt does not shrink the contour in the \(\Omega_\mrm{m}\text{--}\sigma_8\) plane and mildly changes the contour in the \(w_0\text{--}w_a\) plane. The primary contribution from the KL-\(2\times2\)pt is breaking the parameter degeneracy in \(\Omega_\mrm{b}\text{--}h_0\) and \(\Omega_\mrm{m}\text{--}h_0\).\label{fig:KL_2x2_3x2}}
\end{figure*}

\subsection{Limber Regime: Standard \texorpdfstring{$3\times2$pt}{PDFstring} strategy with a spectroscopic sample\label{sec:3x2}}

We use the KL sample for both lens and source galaxy samples in our KL-\(3\times2\)pt analysis. This lens-equal-source idea has been discussed to eliminate nuisance parameters, especially those related to redshift uncertainties, and to improve self-calibration of other galaxy sample-related uncertainties \citep{Fang2022_lenseqsrc, Schaan2020:photoz_self_calib, Boruah2024_lenseqsrc}. For a KL-\(3\times2\)pt analysis, we note that using the full 3D clustering sample may be more beneficial, but we postpone a corresponding analysis to future work. We adopt a scale cut of \(R_\mrm{\min}=21 \,\mrm{Mpc}/h\) for the linear galaxy bias modeling following \citep{Eifler2021_Roman_multiprobe, DESC_SRD}. \tab{\ref{tab:prior}} summarizes the prior ranges and the fiducial values for all parameters. 

We first run a baseline KL-$3\times2$pt analysis with 10 tomographic bins to demonstrate how the constraining power changes as we include more probes. 

In \fg{\ref{fig:KL_2x2_3x2}}, we see a mild increase in constraining power in $\Omega_\mrm{m}\text{--}\sigma_8$ and $w_0\text{--}w_a$ when we combine KL-cosmic shear and KL-$2\times2$pt (the combination of galaxy-galaxy lensing and galaxy clustering). KL-$2\times2$pt alone gives rather loose constraints on these parameters; however, it breaks degeneracies \(\Omega_\mrm{b}\text{--}h_0\) and \(\Omega_\mrm{m}\text{--}h_0\). \(\mrm{FoM}_{\Omega_\mrm{m}\sigma_8}\) changes by 1.4 times and \(\mrm{FoM}_{w_0w_a}\) increases by 1.5 times, as shown in \tab{\ref{tab:LCDM_1D}} and~\ref{tab:wCDM_1D}. 

\begin{figure}
    \centering
    \includegraphics[width=0.99\linewidth]{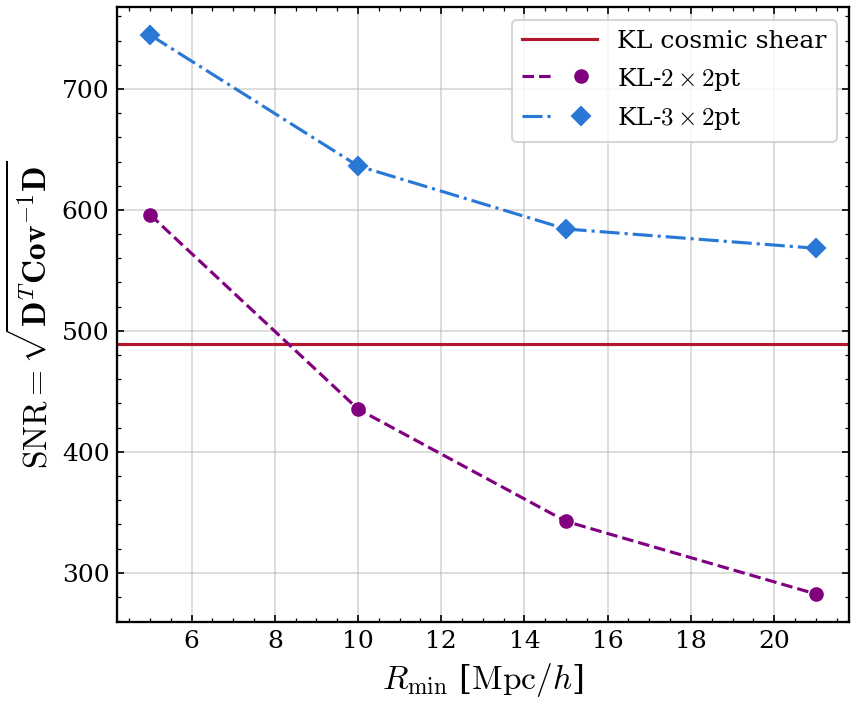}
    \caption{The data vector SNR for KL, KL-\(2\times2\)pt, and KL-\(3\times2\)pt with respect to \(R_{\min}\) in the lens sample. The fiducial scale cut used in this work is 21~\(\mrm{Mpc}/h\).\label{fig:snr}}
\end{figure}

This mild improvement is due to the conservative choice of \(R_{\min}\), which again indicates opportunities for increased constraining power through better modeling of small-scale systematics, in this case galaxy bias. In \fg{\ref{fig:snr}}, we calculate the data vector signal-to-noise ratio (SNR), defined as 
\begin{equation}
    \mathrm{SNR} = \sqrt{\mathbf{D}^{T}\mathbf{Cov}^{-1}\mathbf{D}},
\end{equation}
for KL, KL-\(2\times2\)pt, and KL-\(3\times2\)pt with respect to different \(R_{\min}\). We see that the KL-\(2\times2\)pt SNR decreases rapidly with increasing \(R_{\min}\) and it matches KL-cosmic shear near \(8~\mrm{Mpc}/h\). This analysis motivates significant investment into small-scale modeling of galaxy bias and corresponding priors.

Replacing the KL sample with the main Roman spectroscopic sample for the lens sample to increase number density is another promising avenue to increase constraining power. At lower redshift, the lens sample could be amended by DESI galaxies, which could be combined with the source sample from the Medium Tier. The authors in \citep{Li2026_KL_lensing_ratio} forecast that the constraining power on the dark energy equation of state can be improved by 192\% after including KL shear ratio (KiLeR) in the combined datasets of CMB, BAO, and SN.

\subsection{Non-Limber Regime: Narrow tomographic bins\label{sec:NL}}
Since all galaxies in the KL-\(3\times2\)pt have spectroscopic redshifts, adopting more tomographic bins is an interesting prospect. Such a narrow tomographic-bin analysis would more sensitively probe the time evolution of the dark energy equation of state. We explore this by increasing the number of tomographic bins on a fixed redshift range. 

\begin{figure}
    \centering
    \includegraphics[width=0.99\linewidth]{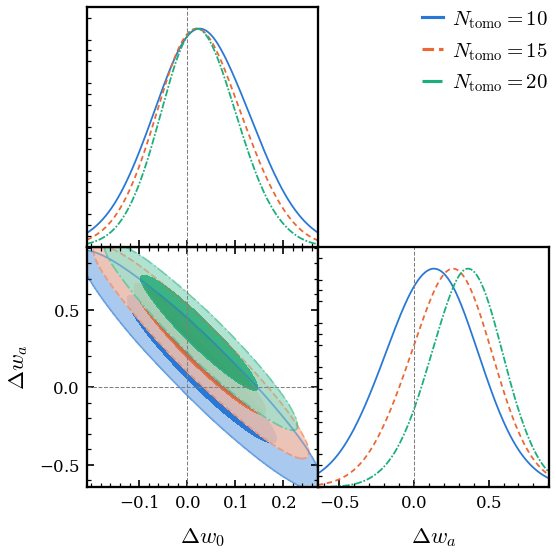} 
    \caption{Comparison of KL-\(3\times2\)pt of 10/15/20 tomographic bins (blue/orange/green) in \(w_0\text{--}w_a\) in \wcdm{} under a misspecification between the data vector and the model. We generate the data vector with non-Limber clustering, while the model uses the Limber approximation. The posterior shifts further away from the fiducial values when the number of tomographic bins increases.\label{fig:mismatch}}
\end{figure}

We examine three scenarios for KL-\(3\times2\)pt: 10 (the baseline), 15, and 20 tomographic bins. The simulated analyses show a mild improvement in \(\mrm{FoM}_{w_0w_a}\) from 34.58 to 45.93 as we increase the number of tomographic bins from 10 to 20. A similar trend appears in \citep{Xu2023} in the cosmic shear-only scenario: increasing the number of tomographic bins does not drastically improve constraining power when the dark energy equation of state evolves slowly over time. 

An additional complication of the narrow-bin KL-\(3\times2\)pt concept is the breakdown of the Limber approximation in the clustering part of the data vector. If non-Limber effects are ignored, this can cause significant parameter biases in the inference \citep{Fang2020, Zhang2026:Limber}. We study this scenario in \fg{\ref{fig:mismatch}}, where we analyze a non-Limber-computed data vector with the Limber model. To isolate the effect of model misspecification, we exclude baryonic effects in these analyses, so the bias shift is stronger in \fg{\ref{fig:mismatch}} than in reality. 

\begin{figure*}
    \centering
    \includegraphics[width=0.9\textwidth]{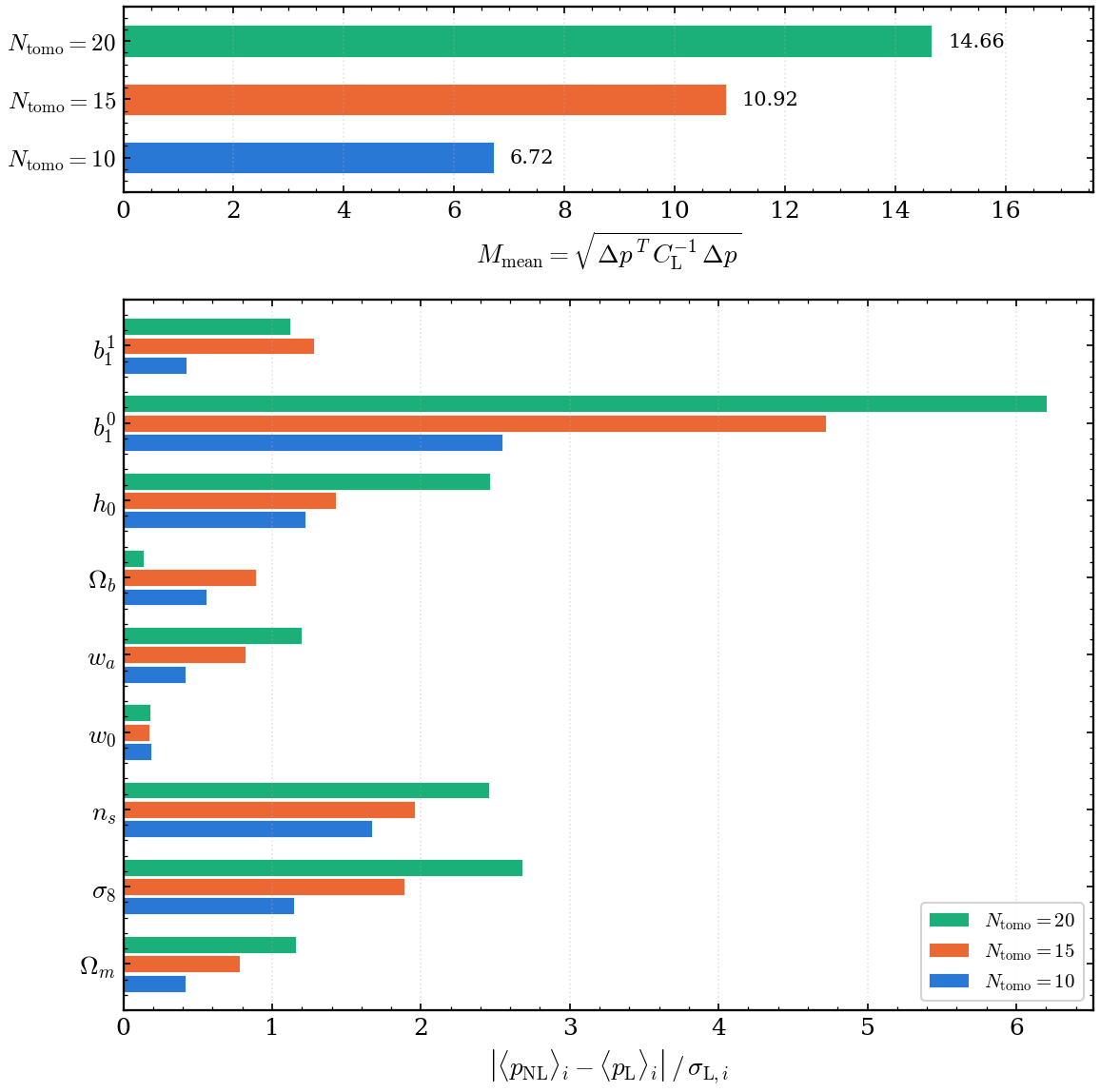}
    \caption{Comparison of bias in each cosmological parameter across tomographic bins in the misspecified scenario. \textit{Top}: the Mahalanobis distance between the Limber and the misspecified posterior means. \textit{Bottom}: the absolute parameter shift between the Limber and the misspecified marginalized posterior mean. These analyses assume \wcdm{} cosmology without baryonic effects to isolate the parameter shift from the model misspecification.\label{fig:bias}}
\end{figure*}

We quantify the parameter shift normalized to the Limber posterior in \fg{\ref{fig:bias}}. The upper panel shows the Mahalanobis distance of the posterior mean shift in the \wcdm{} cosmology \citep{Mahalanobis, Fang2020}. The bottom panel shows the bias in each parameter. The model misspecification impacts \(b_1^0\) and \(\sigma_8\) the most, shifting the 1D marginalized mean by 6.2$\sigma$ and 2.7$\sigma$ in the 20 tomographic bin case. 

\section{Conclusion\label{sec:conclusion}}
In this paper, we perform simulated analyses to forecast KL individually and KL multiprobe strategies with the updated Roman HLWAS survey design. We estimate realistic KL samples using the COSMOS Mock Catalog and derive redshift distributions for the Deep and Medium Tiers of the HLWAS survey. We also consider how telescope roll-angle separation affects KL measurements and explore different scenarios to assess which separation best constrains KL. We compare KL to the traditional photometric WL analysis. Our forecasts include realistic systematic effects, including photometric-redshift and shape-measurement uncertainties, intrinsic alignment, baryonic physics, and galaxy bias.

We find that KL outperforms WL by a factor of $\geq$ two in the FoM for either \lcdm{} or \wcdm. For Year 1 data with 700~deg\(^2\) of full depth, KL alone provides constraining power comparable to the Y5 WL forecast. We note that this study does not include the WL Wide Tier. We further find that Roman's current roll-angle strategy is near-optimal for KL. 

We extend our analysis to explore a multiprobe idea by using the KL sample directly as the lens sample to perform a KL-\(3\times2\)pt analysis. Using the KL sample eliminates photometric uncertainties and enables narrow tomographic cross-correlations between lensing and clustering. Joint clustering and lensing analyses using finer tomographic bins probe the redshift evolution of structure growth and geometry of the universe. While this can better constrain time-dependent dark energy models, the improvement depends strongly on the model's redshift variability; for smooth dark energy evolution, the information gain is not significant. We further show that such an analysis needs to account for non-Limber effects, as they can otherwise bias results in the narrow-tomographic-bin scenario. 

Given KL's low shape noise and corresponding sensitivity to small scales, its performance is more sensitive to baryonic systematics, elevating baryonic systematics to an even greater concern than in standard WL analyses. We quantify the difference in KL constraining power between state-of-the-art baryonic systematics modeling and a scenario where baryonic physics errors are subdominant. We find that there are significant gains possible if baryonic physics models can be improved (either via simulations or self-calibration).

Several possible future extensions of this work are worth noting. First, using the full Roman clustering sample, without downselecting to the KL sample, is an interesting avenue, especially when combining full 3D clustering constraints with KL. This clustering sample can be extended to lower redshift by using DESI or Euclid galaxies. Second, combining WL and KL in a multiprobe analysis is promising, since the WL sample extends to higher redshift. Third, given Roman's projected lifetime of $\sim 22$ years\footnote{\href{https://science.nasa.gov/blogs/roman/2026/09/14/fuel-savings-double-potential-lifetime-for-nasas-roman-mission/}{Fuel Savings Double Potential Lifetime for NASA’s Roman Mission}}, the community could consider extending the current spectroscopic survey to the Roman wide area. While this adds challenges, since the Roman Wide Tier is covered only in H-band \citep{ROTAC2025}, it will strongly boost KL constraining power, given that KL is cosmic-variance limited on most scales.

\begin{acknowledgments} 
This work was supported by NASA Roman WFS 22-ROMAN22-0016. This material is based upon High Performance Computing (HPC) resources supported by the University of Arizona TRIF, UITS, and the Office of Research and Partnerships and maintained by the UArizona Research Technologies department. YHH thanks Haley Bowden for the useful discussion on the WL sample in CPIP DC1.
\end{acknowledgments}

\appendix

\section{Impacts of Roll Angle Separation\label{sec:append_FIM}}
\begin{figure*}
    \centering
    \includegraphics[width=0.49\linewidth]{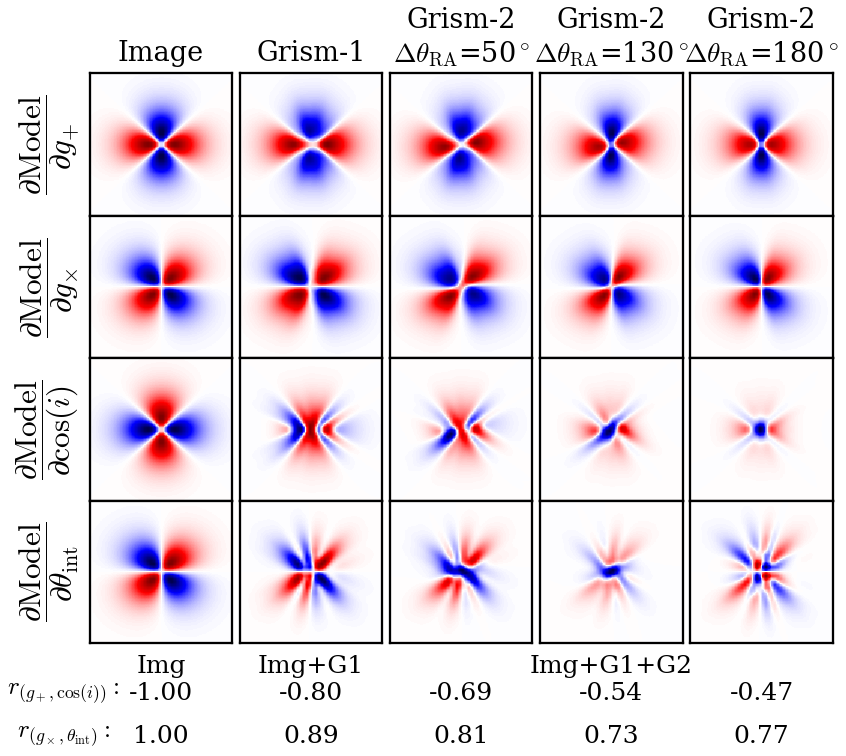}
    \includegraphics[width=0.49\linewidth]{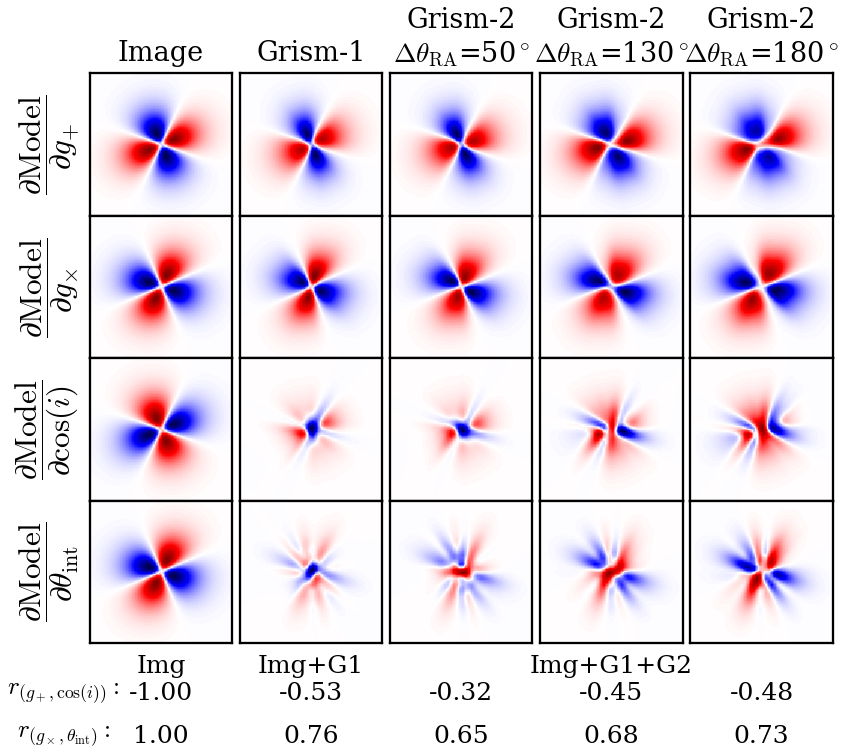}
    \caption{Illustration of the impact of different roll angle separations. Each column shows the partial derivative of the image or grism model with respect to the model parameters. From left to right are the responses of the broadband image, the first grism, the second grism with $50\deg$, $130\deg$, and $180\deg$ roll angle differences $\Delta\theta_\mathrm{RA}$. Each row, from top to bottom, shows the model response to $g_+$, $g_\times$, $\cos i$, and $\theta_\mathrm{int}$. The color map is symmetric about zero, with red (blue) meaning positive (negative) response. We also show the correlation coefficient between degenerate parameter pairs $(g_+,\,\cos i)$ and $(g_\times,\,\theta_\mathrm{int})$ in the bottom for combinations of image-only, image + grism-1, and image + grism-1 + grism-2. The left panel shows an example with $\theta_\mathrm{int}=0\deg$ and the right panel shows when $\theta_\mathrm{int}=280\deg$.}
    \label{fig:append_roll_angle_fisher}
\end{figure*}

In this section, we develop the intuition for how the shear-shape degeneracy breaks when 2D grism spectra are included, and how different roll angles impact the performance. Here we show two examples with different intrinsic position angles in Fig.~\ref{fig:append_roll_angle_fisher} ($\theta_\mathrm{int}=0\deg$ in the left panel, and $\theta_\mathrm{int}=280\deg$ in the right panel). 
The figures show the partial derivatives of mock observations (broadband images and 2D grism spectra) with respect to different model parameters: $g_+$, $g_\times$, $\cos i$, and $\theta_\mathrm{int}$, with blue (red) meaning negative (positive) partial derivative. The perfect degeneracy between $g_+$-$\cos i$ and $g_\times$-$\theta_\mathrm{int}$ can be visualized immediately from the same patterns in the first column of Fig.~\ref{fig:append_roll_angle_fisher}. While the response of 2D grism spectra to $g_+$ and $g_\times$ is very similar to their image response counterparts, the grism response to $\cos i$ and $\theta_\mathrm{int}$ is very different in this case (low inclination $\cos i=0.9$), and adding the first grism observation helps in breaking the degeneracy. When the second grism spectrum is further included, while its responses to $g_+$ and $g_\times$ are still very similar to the image counterparts and change slowly at different roll angle separations, the responses to $\cos i$ and $\theta_\mathrm{int}$ show richer phenomena as a function of roll angle separation. The roll-angle optimization question then turns into which roll angle provides the most independent $\cos i$ and $\theta_\mathrm{int}$ response patterns, and the answer depends on the exact geometry configuration of the galaxy. For the $\theta_\mathrm{int}=0\deg$ example, $\Delta\theta_\mathrm{RA}=180\deg$ provides the most independent $\cos i$ response pattern while $\Delta\theta_\mathrm{RA}=130\deg$ provides the most independent $\theta_\mathrm{int}$ response pattern among the three roll angle examples. The $\theta_\mathrm{int}=280\deg$ example has a different preference.

\twocolumngrid 
\bibliography{main}

@PREAMBLE{
"\providecommand{\mnras}{Monthly Notices of the Royal Astronomical Society}"
# "\providecommand{\araa}{Annual Review of Astronomy and Astrophysics}"
# "\providecommand{\apjs}{The Astrophysical Journal Supplement}"
# "\providecommand{\apj}{The Astrophysical Journal}"
# "\providecommand{\apjl}{The Astrophysical Journal Letter}"
# "\providecommand{\aap}{Astronomy and Astrophysics}"
# "\providecommand{\physrep}{Physics Reports}"
# "\providecommand{\pasj}{Publications of the ASJ}"
# "\providecommand{\aj}{The Astronomical Journal}"
# "\providecommand{\pasa}{Publications of the Astronomical Society of Astralia}"
# "\providecommand{\pasp}{Publications of the Astronomical Society of the Pacific}"
# "\providecommand{\jcap}{Journal of Cosmology and Astroparticle Physics}"
# "\providecommand{\pasj}{Publications of the Astronomical Society of Japan}"
}

@ARTICLE{WFIRST_AFTA,
       author = {{Spergel}, D. and {Gehrels}, N. and {Baltay}, C. and {Bennett}, D. and {Breckinridge}, J. and {Donahue}, M. and {Dressler}, A. and {Gaudi}, B.~S. and {Greene}, T. and {Guyon}, O. and {Hirata}, C. and {Kalirai}, J. and {Kasdin}, N.~J. and {Macintosh}, B. and {Moos}, W. and {Perlmutter}, S. and {Postman}, M. and {Rauscher}, B. and {Rhodes}, J. and {Wang}, Y. and {Weinberg}, D. and {Benford}, D. and {Hudson}, M. and {Jeong}, W.-S. and {Mellier}, Y. and {Traub}, W. and {Yamada}, T. and {Capak}, P. and {Colbert}, J. and {Masters}, D. and {Penny}, M. and {Savransky}, D. and {Stern}, D. and {Zimmerman}, N. and {Barry}, R. and {Bartusek}, L. and {Carpenter}, K. and {Cheng}, E. and {Content}, D. and {Dekens}, F. and {Demers}, R. and {Grady}, K. and {Jackson}, C. and {Kuan}, G. and {Kruk}, J. and {Melton}, M. and {Nemati}, B. and {Parvin}, B. and {Poberezhskiy}, I. and {Peddie}, C. and {Ruffa}, J. and {Wallace}, J.~K. and {Whipple}, A. and {Wollack}, E. and {Zhao}, F.},
        title = "{Wide-Field InfrarRed Survey Telescope-Astrophysics Focused Telescope Assets WFIRST-AFTA 2015 Report}",
      journal = {arXiv e-prints},
         year = 2015,
        month = mar,
          eid = {arXiv:1503.03757},
        pages = {arXiv:1503.03757},
          doi = {10.48550/arXiv.1503.03757},
archivePrefix = {arXiv},
       eprint = {1503.03757},
 primaryClass = {astro-ph.IM},
       adsurl = {https://ui.adsabs.harvard.edu/abs/2015arXiv150303757S}
}

@ARTICLE{wang_hlss_2022,
       author = {{Wang}, Yun and {Zhai}, Zhongxu and {Alavi}, Anahita and {Massara}, Elena and {Pisani}, Alice and {Benson}, Andrew and {Hirata}, Christopher M. and {Samushia}, Lado and {Weinberg}, David H. and {Colbert}, James and {Dor{\'e}}, Olivier and {Eifler}, Tim and {Heinrich}, Chen and {Ho}, Shirley and {Krause}, Elisabeth and {Padmanabhan}, Nikhil and {Spergel}, David and {Teplitz}, Harry I.},
        title = "{The High Latitude Spectroscopic Survey on the Nancy Grace Roman Space Telescope}",
      journal = {\apj},
         year = 2022,
        month = mar,
       volume = {928},
       number = {1},
          eid = {1},
        pages = {1},
          doi = {10.3847/1538-4357/ac4973},
archivePrefix = {arXiv},
       eprint = {2110.01829},
 primaryClass = {astro-ph.CO},
       adsurl = {https://ui.adsabs.harvard.edu/abs/2022ApJ...928....1W}
}

@ARTICLE{cosmolike,
       author = {{Krause}, Elisabeth and {Eifler}, Tim},
        title = "{cosmolike - cosmological likelihood analyses for photometric galaxy surveys}",
      journal = {\mnras},
         year = 2017,
        month = sep,
       volume = {470},
       number = {2},
        pages = {2100-2112},
          doi = {10.1093/mnras/stx1261},
archivePrefix = {arXiv},
       eprint = {1601.05779},
 primaryClass = {astro-ph.CO},
       adsurl = {https://ui.adsabs.harvard.edu/abs/2017MNRAS.470.2100K}
}

@ARTICLE{Li2026_KL_lensing_ratio,
       author = {{Li}, Qinxun},
        title = "{Kinematic Lensing Ratio: Reviving Weak Lensing Cosmography as a Geometric Dark Energy Probe}",
      journal = {arXiv e-prints},
         year = 2026,
        month = apr,
          eid = {arXiv:2604.26015},
        pages = {arXiv:2604.26015},
          doi = {10.48550/arXiv.2604.26015},
archivePrefix = {arXiv},
       eprint = {2604.26015},
 primaryClass = {astro-ph.CO},
       adsurl = {https://ui.adsabs.harvard.edu/abs/2026arXiv260426015L}
}

@ARTICLE{ROTAC2025,
       author = {Zasowski, Gail and others},
        title = "{Roman Observations Time Allocation Committee: Final Report and Recommendations}",
      journal = {arXiv e-prints},
         year = 2025,
        month = may,
          eid = {arXiv:2505.10574},
        pages = {arXiv:2505.10574},
          doi = {10.48550/arXiv.2505.10574},
archivePrefix = {arXiv},
       eprint = {2505.10574},
 primaryClass = {astro-ph.IM},
       adsurl = {https://ui.adsabs.harvard.edu/abs/2025arXiv250510574O}
}

@ARTICLE{OC20,
       author = {{Outini}, M. and {Copin}, Y.},
        title = "{Forward modeling of galaxy kinematics in slitless spectroscopy}",
      journal = {\aap},
         year = 2020,
        month = jan,
       volume = {633},
          eid = {A43},
        pages = {A43},
          doi = {10.1051/0004-6361/201936318},
archivePrefix = {arXiv},
       eprint = {1910.07803},
 primaryClass = {astro-ph.GA},
       adsurl = {https://ui.adsabs.harvard.edu/abs/2020A&A...633A..43O}
}

@ARTICLE{Jouvel2009,
       author = {{Jouvel}, S. and {Kneib}, J. -P. and {Ilbert}, O. and {Bernstein}, G. and {Arnouts}, S. and {Dahlen}, T. and {Ealet}, A. and {Milliard}, B. and {Aussel}, H. and {Capak}, P. and {Koekemoer}, A. and {Le Brun}, V. and {McCracken}, H. and {Salvato}, M. and {Scoville}, N.},
        title = "{Designing future dark energy space missions. I. Building realistic galaxy spectro-photometric catalogs and their first applications}",
      journal = {\aap},
         year = 2009,
        month = sep,
       volume = {504},
       number = {2},
        pages = {359-371},
          doi = {10.1051/0004-6361/200911798},
archivePrefix = {arXiv},
       eprint = {0902.0625},
 primaryClass = {astro-ph.CO},
       adsurl = {https://ui.adsabs.harvard.edu/abs/2009A&A...504..359J}
}

@ARTICLE{Eifler_LSST_Roman,
       author = {{Eifler}, Tim and {Simet}, Melanie and {Krause}, Elisabeth and {Hirata}, Christopher and {Huang}, Hung-Jin and {Fang}, Xiao and {Miranda}, Vivian and {Mandelbaum}, Rachel and {Doux}, Cyrille and {Heinrich}, Chen and {Huff}, Eric and {Miyatake}, Hironao and {Hemmati}, Shoubaneh and {Xu}, Jiachuan and {Rogozenski}, Paul and {Capak}, Peter and {Choi}, Ami and {Dor{\'e}}, Olivier and {Jain}, Bhuvnesh and {Jarvis}, Mike and {Kruk}, Jeffrey and {MacCrann}, Niall and {Masters}, Dan and {Rozo}, Eduardo and {Spergel}, David N. and {Troxel}, Michael and {von der Linden}, Anja and {Wang}, Yun and {Weinberg}, David H. and {Wenzl}, Lukas and {Wu}, Hao-Yi},
        title = "{Cosmology with the Roman Space Telescope: synergies with the Rubin Observatory Legacy Survey of Space and Time}",
      journal = {\mnras},
         year = 2021,
        month = oct,
       volume = {507},
       number = {1},
        pages = {1514-1527},
          doi = {10.1093/mnras/stab533},
archivePrefix = {arXiv},
       eprint = {2004.04702},
 primaryClass = {astro-ph.CO},
       adsurl = {https://ui.adsabs.harvard.edu/abs/2021MNRAS.507.1514E}
}

@ARTICLE{Fang2020,
       author = {{Fang}, Xiao and {Krause}, Elisabeth and {Eifler}, Tim and {MacCrann}, Niall},
        title = "{Beyond Limber: efficient computation of angular power spectra for galaxy clustering and weak lensing}",
      journal = {\jcap},
         year = 2020,
        month = may,
       volume = {2020},
       number = {5},
          eid = {010},
        pages = {010},
          doi = {10.1088/1475-7516/2020/05/010},
archivePrefix = {arXiv},
       eprint = {1911.11947},
 primaryClass = {astro-ph.CO},
       adsurl = {https://ui.adsabs.harvard.edu/abs/2020JCAP...05..010F}
}

@ARTICLE{Ma2006_photoz,
       author = {{Ma}, Zhaoming and {Hu}, Wayne and {Huterer}, Dragan},
        title = "{Effects of Photometric Redshift Uncertainties on Weak-Lensing Tomography}",
      journal = {\apj},
         year = 2006,
        month = jan,
       volume = {636},
       number = {1},
        pages = {21-29},
          doi = {10.1086/497068},
archivePrefix = {arXiv},
       eprint = {astro-ph/0506614},
 primaryClass = {astro-ph},
       adsurl = {https://ui.adsabs.harvard.edu/abs/2006ApJ...636...21M}
}

@ARTICLE{Krause2016_IA,
       author = {{Krause}, Elisabeth and {Eifler}, Tim and {Blazek}, Jonathan},
        title = "{The impact of intrinsic alignment on current and future cosmic shear surveys}",
      journal = {\mnras},
         year = 2016,
        month = feb,
       volume = {456},
       number = {1},
        pages = {207-222},
          doi = {10.1093/mnras/stv2615},
archivePrefix = {arXiv},
       eprint = {1506.08730},
 primaryClass = {astro-ph.CO},
       adsurl = {https://ui.adsabs.harvard.edu/abs/2016MNRAS.456..207K}
}

@article{Hirata2004_IA,
  title = {Intrinsic alignment-lensing interference as a contaminant of cosmic shear},
  author = {{Hirata}, Christopher M. and {Seljak}, Uro{\v{s}}},
  journal = {Phys. Rev. D},
  volume = {70},
  issue = {6},
  pages = {063526},
  numpages = {11},
  year = {2004},
  month = {Sep},
  publisher = {American Physical Society},
  doi = {10.1103/PhysRevD.70.063526},
  url = {https://link.aps.org/doi/10.1103/PhysRevD.70.063526}
}

@ARTICLE{HJHuang2019_baryon,
       author = {{Huang}, Hung-Jin and {Eifler}, Tim and {Mandelbaum}, Rachel and {Dodelson}, Scott},
        title = "{Modelling baryonic physics in future weak lensing surveys}",
      journal = {\mnras},
         year = 2019,
        month = sep,
       volume = {488},
       number = {2},
        pages = {1652-1678},
          doi = {10.1093/mnras/stz1714},
archivePrefix = {arXiv},
       eprint = {1809.01146},
 primaryClass = {astro-ph.CO},
       adsurl = {https://ui.adsabs.harvard.edu/abs/2019MNRAS.488.1652H}
}

@ARTICLE{Eifler2015_baryon,
       author = {{Eifler}, Tim and {Krause}, Elisabeth and {Dodelson}, Scott and {Zentner}, Andrew R. and {Hearin}, Andrew P. and {Gnedin}, Nickolay Y.},
        title = "{Accounting for baryonic effects in cosmic shear tomography: determining a minimal set of nuisance parameters using PCA}",
      journal = {\mnras},
         year = 2015,
        month = dec,
       volume = {454},
       number = {3},
        pages = {2451-2471},
          doi = {10.1093/mnras/stv2000},
archivePrefix = {arXiv},
       eprint = {1405.7423},
 primaryClass = {astro-ph.CO},
       adsurl = {https://ui.adsabs.harvard.edu/abs/2015MNRAS.454.2451E}
}

@ARTICLE{DESI_DR2_BAO,
       author = {{DESI Collaboration} and {Abdul Karim}, M. and {Aguilar}, J. and {Ahlen}, S. and {Alam}, S. and {Allen}, L. and {Allende Prieto}, C. and {Alves}, O. and {Anand}, A. and {Andrade}, U. and {Armengaud}, E. and et al.},
        title = "{DESI DR2 results. II. Measurements of baryon acoustic oscillations and cosmological constraints}",
      journal = {\prd},
         year = 2025,
        month = oct,
       volume = {112},
       number = {8},
          eid = {083515},
        pages = {083515},
          doi = {10.1103/tr6y-kpc6},
archivePrefix = {arXiv},
       eprint = {2503.14738},
 primaryClass = {astro-ph.CO},
       adsurl = {https://ui.adsabs.harvard.edu/abs/2025PhRvD.112h3515A}
}

@ARTICLE{Lodha2025_DDE,
       author = {{Lodha}, K. and {Calderon}, R. and {Matthewson}, W.~L. and {Shafieloo}, A. and {Ishak}, M. and {Pan}, J. and {Garcia-Quintero}, C. and {Huterer}, D. and {Valogiannis}, G. and {Ure{\~n}a-L{\'o}pez}, L.~A. and {Kamble}, N.~V. and {Parkinson}, D. and {Kim}, A.~G. and {Zhao}, G.~B. and {Cervantes-Cota}, J.~L. and {Rohlf}, J. and {Lozano-Rodr{\'\i}guez}, F. and {Rom{\'a}n-Herrera}, J.~O. and {Abdul-Karim}, M. and {Aguilar}, J. and {Ahlen}, S. and {Alves}, O. and {Andrade}, U. and {Armengaud}, E. and {Aviles}, A. and {Behera}, J. and {BenZvi}, S. and {Bianchi}, D. and {Brodzeller}, A. and {Brooks}, D. and {Burtin}, E. and {Canning}, R. and {Rosell}, A. Carnero and {Casas}, L. and {Castander}, F.~J. and {Charles}, M. and {Chaussidon}, E. and {Chaves-Montero}, J. and {Chebat}, D. and {Claybaugh}, T. and {Cole}, S. and {Cuceu}, A. and {Dawson}, K.~S. and {de la Macorra}, A. and {de Mattia}, A. and {Deiosso}, N. and {Demina}, R. and {Dey}, Arjun and {Dey}, Biprateep and {Ding}, Z. and {Doel}, P. and {Eisenstein}, D.~J. and {Elbers}, W. and {Ferraro}, S. and {Font-Ribera}, A. and {Forero-Romero}, J.~E. and {Garrison}, Lehman H. and {Gazta{\~n}aga}, E. and {Gil-Mar{\'\i}n}, H. and {Gontcho}, S. Gontcho A. and {Gonzalez-Morales}, A.~X. and {Gutierrez}, G. and {Guy}, J. and {Hahn}, C. and {Herbold}, M. and {Herrera-Alcantar}, H.~K. and {Honscheid}, K. and {Howlett}, C. and {Juneau}, S. and {Kehoe}, R. and {Kirkby}, D. and {Kisner}, T. and {Kremin}, A. and {Lahav}, O. and {Lamman}, C. and {Landriau}, M. and {Le Guillou}, L. and {Leauthaud}, A. and {Levi}, M.~E. and {Li}, Q. and {Magneville}, C. and {Manera}, M. and {Martini}, P. and {Meisner}, A. and {Mena-Fern{\'a}ndez}, J. and {Miquel}, R. and {Moustakas}, J. and {Santos}, D. Mu{\~n}oz and {Mu{\~n}oz-Guti{\'e}rrez}, A. and {Myers}, A.~D. and {Nadathur}, S. and {Niz}, G. and {Noriega}, H.~E. and {Paillas}, E. and {Palanque-Delabrouille}, N. and {Percival}, W.~J. and {Pieri}, Matthew M. and {Poppett}, C. and {Prada}, F. and {P{\'e}rez-Fern{\'a}ndez}, A. and {P{\'e}rez-R{\`a}fols}, I. and {Ram{\'\i}rez-P{\'e}rez}, C. and {Rashkovetskyi}, M. and {Ravoux}, C. and {Ross}, A.~J. and {Rossi}, G. and {Ruhlmann-Kleider}, V. and {Samushia}, L. and {Sanchez}, E. and {Schlegel}, D. and {Schubnell}, M. and {Seo}, H. and {Sinigaglia}, F. and {Sprayberry}, D. and {Tan}, T. and {Tarl{\'e}}, G. and {Taylor}, P. and {Turner}, W. and {Vargas-Maga{\~n}a}, M. and {Walther}, M. and {Weaver}, B.~A. and {Wolfson}, M. and {Y{\`e}che}, C. and {Zarrouk}, P. and {Zhou}, R. and {Zou}, H. and {DESI Collaboration}},
        title = "{Extended dark energy analysis using DESI DR2 BAO measurements}",
      journal = {\prd},
         year = 2025,
        month = oct,
       volume = {112},
       number = {8},
          eid = {083511},
        pages = {083511},
          doi = {10.1103/w4c6-1r5j},
archivePrefix = {arXiv},
       eprint = {2503.14743},
 primaryClass = {astro-ph.CO},
       adsurl = {https://ui.adsabs.harvard.edu/abs/2025PhRvD.112h3511L}
}

@ARTICLE{Corte_Liddle2025_noSN,
       author = {{Cort{\^e}s}, Marina and {Liddle}, Andrew R.},
        title = "{On DESI's DR2 exclusion of {\ensuremath{\Lambda}}CDM}",
      journal = {\mnras},
         year = 2025,
        month = nov,
       volume = {544},
       number = {1},
        pages = {L121-L125},
          doi = {10.1093/mnrasl/slaf108},
archivePrefix = {arXiv},
       eprint = {2504.15336},
 primaryClass = {astro-ph.CO},
       adsurl = {https://ui.adsabs.harvard.edu/abs/2025MNRAS.544L.121C}
}

@ARTICLE{Popovic2026_SNsys,
       author = {{Popovic}, B. and {Shah}, P. and {Kenworthy}, W.~D. and {Kessler}, R. and {Davis}, T.~M. and {Goobar}, A. and {Scolnic}, D. and {Vincenzi}, M. and {Wiseman}, P. and {Chen}, R. and {Charleton}, E. and {Acevedo}, M. and {Armstrong}, P. and {Boyd}, B.~M. and {Brout}, D. and {Camilleri}, R. and {Frieman}, J. and {Galbany}, L. and {Grayling}, M. and {Kelsey}, L. and {Rose}, B. and {S{\'a}nchez}, B. and {Lee}, J. and {M{\"o}ller}, A. and {Smith}, M. and {Sullivan}, M. and {Shiamtanis}, N. and {Alarcon}, A. and {Allam}, S.~S. and {Andrade-Oliveira}, F. and {Avila}, S. and {Bacon}, D. and {Blazek}, J. and {Bocquet}, S. and {Brooks}, D. and {Burke}, D.~L. and {Rosell}, A. Carnero and {Carretero}, J. and {Cawthon}, R. and {da Costa}, L.~N. and {Pereira}, M.~E. da Silva and {Diehl}, H.~T. and {Dodelson}, S. and {Doel}, P. and {Everett}, S. and {Frohmaier}, C. and {Garc{\'\i}a-Bellido}, J. and {Gruen}, D. and {Gutierrez}, G. and {Herner}, K. and {Hinton}, S.~R. and {Hollowood}, D.~L. and {Honscheid}, K. and {Huterer}, D. and {James}, D.~J. and {Jeffrey}, N. and {Kuehn}, K. and {Lahav}, O. and {Lee}, S. and {Lidman}, C. and {Marshall}, J.~L. and {Mena-Fern{\'a}ndez}, J. and {Menanteau}, F. and {Miquel}, R. and {Muir}, J. and {Myles}, J. and {Ogando}, R.~L.~C. and {Paterno}, M. and {Malag{\'o}n}, A.~A. Plazas and {Porredon}, A. and {Prat}, J. and {Nichol}, R.~C. and {Romer}, A.~K. and {Roodman}, A. and {Sanchez}, E. and {Cid}, D. Sanchez and {Sevilla-Noarbe}, I. and {Suchyta}, E. and {Swanson}, M.~E.~C. and {To}, C. and {Tucker}, D.~L. and {Walker}, A.~R. and {Weaverdyck}, N. and {Aguena}, M.},
        title = "{The Dark Energy Survey supernova program: a reanalysis of cosmology results and evidence for evolving dark energy with an updated Type Ia supernova calibration}",
      journal = {\mnras},
         year = 2026,
        month = jun,
       volume = {548},
       number = {4},
          eid = {stag632},
        pages = {stag632},
          doi = {10.1093/mnras/stag632},
archivePrefix = {arXiv},
       eprint = {2511.07517},
 primaryClass = {astro-ph.CO},
       adsurl = {https://ui.adsabs.harvard.edu/abs/2026MNRAS.548ag632P}
}

@ARTICLE{Efstathiou2025_degeneracy,
       author = {{Efstathiou}, George},
        title = "{Baryon acoustic oscillations from a different angle}",
      journal = {\mnras},
         year = 2025,
        month = jul,
       volume = {540},
       number = {3},
        pages = {2844-2852},
          doi = {10.1093/mnras/staf906},
archivePrefix = {arXiv},
       eprint = {2505.02658},
 primaryClass = {astro-ph.CO},
       adsurl = {https://ui.adsabs.harvard.edu/abs/2025MNRAS.540.2844E}
}

@ARTICLE{Efstathiou2025_SNsys,
       author = {{Efstathiou}, George},
        title = "{Evolving dark energy or supernovae systematics?}",
      journal = {\mnras},
         year = 2025,
        month = apr,
       volume = {538},
       number = {2},
        pages = {875-882},
          doi = {10.1093/mnras/staf301},
archivePrefix = {arXiv},
       eprint = {2408.07175},
 primaryClass = {astro-ph.CO},
       adsurl = {https://ui.adsabs.harvard.edu/abs/2025MNRAS.538..875E}
}

@article{Euclid2024_overview,
    author = "{Euclid Collaboration}",
    title = "{Euclid - I. Overview of the Euclid mission}",
    eprint = "2405.13491",
    archivePrefix = "arXiv",
    primaryClass = "astro-ph.CO",
    doi = "10.1051/0004-6361/202450810",
    journal = "Astron. Astrophys.",
    volume = "697",
    pages = "A1",
    year = "2025"
}

@ARTICLE{TED,
       author = {{Huang}, Yu-Hsiu and {Krause}, Elisabeth and {Xu}, Jiachuan and {Eifler}, Tim and {R.~S.}, Pranjal and {Huff}, Eric},
        title = "{Astrophysical systematics in kinematic lensing: Quantifying an intrinsic alignment analog}",
      journal = {\prd},
         year = 2024,
        month = aug,
       volume = {110},
       number = {4},
          eid = {043509},
        pages = {043509},
          doi = {10.1103/PhysRevD.110.043509},
archivePrefix = {arXiv},
       eprint = {2404.00197},
 primaryClass = {astro-ph.CO},
       adsurl = {https://ui.adsabs.harvard.edu/abs/2024PhRvD.110d3509H}
}

@ARTICLE{Xu2023,
       author = {{Xu}, Jiachuan and {Eifler}, Tim and {Huff}, Eric and {Pranjal}, R.~S. and {Huang}, Hung-Jin and {Everett}, Spencer and {Krause}, Elisabeth},
        title = "{Kinematic lensing with the Roman Space Telescope}",
      journal = {\mnras},
         year = 2023,
        month = feb,
       volume = {519},
       number = {2},
        pages = {2535-2551},
          doi = {10.1093/mnras/stac3685},
archivePrefix = {arXiv},
       eprint = {2201.00739},
 primaryClass = {astro-ph.CO},
       adsurl = {https://ui.adsabs.harvard.edu/abs/2023MNRAS.519.2535X}
}

@ARTICLE{RS2025_pilot,
       author = {{R.~S.}, Pranjal and {Huff}, Eric and {Krause}, Elisabeth and {Eifler}, Tim and {Everett}, Spencer and {Huang}, Yu-Hsiu and {Xu}, Jiachuan},
        title = "{Kinematic lensing inference - II. Cluster lensing with $\mathcal{O}(1)$ galaxies}",
      journal = {\mnras},
         year = 2025,
        month = jul,
       volume = {540},
       number = {4},
        pages = {2877-2893},
          doi = {10.1093/mnras/staf845},
archivePrefix = {arXiv},
       eprint = {2409.08367},
 primaryClass = {astro-ph.CO},
       adsurl = {https://ui.adsabs.harvard.edu/abs/2025MNRAS.540.2877P}
}

@ARTICLE{RS2023_shape_noise,
       author = {{R.~S.}, Pranjal and {Krause}, Elisabeth and {Huang}, Hung-Jin and {Huff}, Eric and {Xu}, Jiachuan and {Eifler}, Tim and {Everett}, Spencer},
        title = "{Kinematic lensing inference - I. Characterizing shape noise with simulated analyses}",
      journal = {\mnras},
         year = 2023,
        month = sep,
       volume = {524},
       number = {3},
        pages = {3324-3334},
          doi = {10.1093/mnras/stad2014},
archivePrefix = {arXiv},
       eprint = {2209.11811},
 primaryClass = {astro-ph.GA},
       adsurl = {https://ui.adsabs.harvard.edu/abs/2023MNRAS.524.3324P}
}

@ARTICLE{Huff2013_KL,
       author = {{Huff}, Eric M. and {Krause}, Elisabeth and {Eifler}, Tim and {Fang}, Xiao and {George}, Matthew R. and {Schlegel}, David},
        title = "{Cosmic shear without shape noise}",
      journal = {arXiv e-prints},
         year = 2013,
        month = nov,
          eid = {arXiv:1311.1489},
        pages = {arXiv:1311.1489},
          doi = {10.48550/arXiv.1311.1489},
archivePrefix = {arXiv},
       eprint = {1311.1489},
 primaryClass = {astro-ph.CO},
       adsurl = {https://ui.adsabs.harvard.edu/abs/2013arXiv1311.1489H}
}

@ARTICLE{Xu2024,
       author = {{Xu}, Jiachuan and {Eifler}, Tim and {Wang}, Eason and {Krause}, Elisabeth and {Everett}, Spencer and {Huff}, Eric and {Pranjal R.}, S. and {Huang}, Yu-Hsiu},
        title = "{Kinematic Lensing with the Dark Energy Spectroscopic Instrument -- Probing structure formation at very low redshift}",
      journal = {arXiv e-prints},
         year = 2024,
        month = jul,
          eid = {arXiv:2407.20867},
        pages = {arXiv:2407.20867},
          doi = {10.48550/arXiv.2407.20867},
archivePrefix = {arXiv},
       eprint = {2407.20867},
 primaryClass = {astro-ph.CO},
       adsurl = {https://ui.adsabs.harvard.edu/abs/2024arXiv240720867X}
}

@ARTICLE{Huang2026,
       author = {{Huang}, Yu-Hsiu and {Krause}, Elisabeth and {Eifler}, Tim and {Bernstein}, Gary and {Xu}, Jiachuan and {Huff}, Eric and {Pranjal R.~S.}},
        title = "{Cosmic shear with one component and its application to future radio surveys}",
      journal = {\prd},
         year = 2026,
        month = may,
       volume = {113},
       number = {10},
          eid = {103534},
        pages = {103534},
          doi = {10.1103/bfx7-4fsw},
archivePrefix = {arXiv},
       eprint = {2510.18011},
 primaryClass = {astro-ph.CO},
       adsurl = {https://ui.adsabs.harvard.edu/abs/2026PhRvD.113j3534H}
}

@ARTICLE{Eifler2021_Roman_multiprobe,
       author = {{Eifler}, Tim and {Miyatake}, Hironao and {Krause}, Elisabeth and {Heinrich}, Chen and {Miranda}, Vivian and {Hirata}, Christopher and {Xu}, Jiachuan and {Hemmati}, Shoubaneh and {Simet}, Melanie and {Capak}, Peter and {Choi}, Ami and {Dor{\'e}}, Olivier and {Doux}, Cyrille and {Fang}, Xiao and {Hounsell}, Rebekah and {Huff}, Eric and {Huang}, Hung-Jin and {Jarvis}, Mike and {Kruk}, Jeffrey and {Masters}, Dan and {Rozo}, Eduardo and {Scolnic}, Dan and {Spergel}, David N. and {Troxel}, Michael and {von der Linden}, Anja and {Wang}, Yun and {Weinberg}, David H. and {Wenzl}, Lukas and {Wu}, Hao-Yi},
        title = "{Cosmology with the Roman Space Telescope - multiprobe strategies}",
      journal = {\mnras},
         year = 2021,
        month = oct,
       volume = {507},
       number = {2},
        pages = {1746-1761},
          doi = {10.1093/mnras/stab1762},
archivePrefix = {arXiv},
       eprint = {2004.05271},
 primaryClass = {astro-ph.CO},
       adsurl = {https://ui.adsabs.harvard.edu/abs/2021MNRAS.507.1746E}
}

@ARTICLE{Grogin2011_CANDELS,
       author = {{Grogin}, Norman A. and {Kocevski}, Dale D. and {Faber}, S.~M. and {Ferguson}, Henry C. and {Koekemoer}, Anton M. and {Riess}, Adam G. and {Acquaviva}, Viviana and {Alexander}, David M. and {Almaini}, Omar and {Ashby}, Matthew L.~N. and {Barden}, Marco and {Bell}, Eric F. and {Bournaud}, Fr{\'e}d{\'e}ric and {Brown}, Thomas M. and {Caputi}, Karina I. and {Casertano}, Stefano and {Cassata}, Paolo and {Castellano}, Marco and {Challis}, Peter and {Chary}, Ranga-Ram and {Cheung}, Edmond and {Cirasuolo}, Michele and {Conselice}, Christopher J. and {Roshan Cooray}, Asantha and {Croton}, Darren J. and {Daddi}, Emanuele and {Dahlen}, Tomas and {Dav{\'e}}, Romeel and {de Mello}, Du{\'\i}lia F. and {Dekel}, Avishai and {Dickinson}, Mark and {Dolch}, Timothy and {Donley}, Jennifer L. and {Dunlop}, James S. and {Dutton}, Aaron A. and {Elbaz}, David and {Fazio}, Giovanni G. and {Filippenko}, Alexei V. and {Finkelstein}, Steven L. and {Fontana}, Adriano and {Gardner}, Jonathan P. and {Garnavich}, Peter M. and {Gawiser}, Eric and {Giavalisco}, Mauro and {Grazian}, Andrea and {Guo}, Yicheng and {Hathi}, Nimish P. and {H{\"a}ussler}, Boris and {Hopkins}, Philip F. and {Huang}, Jia-Sheng and {Huang}, Kuang-Han and {Jha}, Saurabh W. and {Kartaltepe}, Jeyhan S. and {Kirshner}, Robert P. and {Koo}, David C. and {Lai}, Kamson and {Lee}, Kyoung-Soo and {Li}, Weidong and {Lotz}, Jennifer M. and {Lucas}, Ray A. and {Madau}, Piero and {McCarthy}, Patrick J. and {McGrath}, Elizabeth J. and {McIntosh}, Daniel H. and {McLure}, Ross J. and {Mobasher}, Bahram and {Moustakas}, Leonidas A. and {Mozena}, Mark and {Nandra}, Kirpal and {Newman}, Jeffrey A. and {Niemi}, Sami-Matias and {Noeske}, Kai G. and {Papovich}, Casey J. and {Pentericci}, Laura and {Pope}, Alexandra and {Primack}, Joel R. and {Rajan}, Abhijith and {Ravindranath}, Swara and {Reddy}, Naveen A. and {Renzini}, Alvio and {Rix}, Hans-Walter and {Robaina}, Aday R. and {Rodney}, Steven A. and {Rosario}, David J. and {Rosati}, Piero and {Salimbeni}, Sara and {Scarlata}, Claudia and {Siana}, Brian and {Simard}, Luc and {Smidt}, Joseph and {Somerville}, Rachel S. and {Spinrad}, Hyron and {Straughn}, Amber N. and {Strolger}, Louis-Gregory and {Telford}, Olivia and {Teplitz}, Harry I. and {Trump}, Jonathan R. and {van der Wel}, Arjen and {Villforth}, Carolin and {Wechsler}, Risa H. and {Weiner}, Benjamin J. and {Wiklind}, Tommy and {Wild}, Vivienne and {Wilson}, Grant and {Wuyts}, Stijn and {Yan}, Hao-Jing and {Yun}, Min S.},
        title = "{CANDELS: The Cosmic Assembly Near-infrared Deep Extragalactic Legacy Survey}",
      journal = {\apjs},
         year = 2011,
        month = dec,
       volume = {197},
       number = {2},
          eid = {35},
        pages = {35},
          doi = {10.1088/0067-0049/197/2/35},
archivePrefix = {arXiv},
       eprint = {1105.3753},
 primaryClass = {astro-ph.CO},
       adsurl = {https://ui.adsabs.harvard.edu/abs/2011ApJS..197...35G}
}

@ARTICLE{Hirata2012_ETC,
       author = {{Hirata}, Christopher M. and {Gehrels}, Neil and {Kneib}, Jean-Paul and {Kruk}, Jeffrey and {Rhodes}, Jason and {Wang}, Yun and {Zoubian}, Julien},
        title = "{The WFIRST Galaxy Survey Exposure Time Calculator}",
      journal = {arXiv e-prints},
         year = 2012,
        month = apr,
          eid = {arXiv:1204.5151},
        pages = {arXiv:1204.5151},
          doi = {10.48550/arXiv.1204.5151},
archivePrefix = {arXiv},
       eprint = {1204.5151},
 primaryClass = {astro-ph.IM},
       adsurl = {https://ui.adsabs.harvard.edu/abs/2012arXiv1204.5151H}
}

@ARTICLE{DES2026:DE,
       author = {{DES Collaboration} and {Abbott}, T.~M.~C. and {Adamow}, M. and {Aguena}, M. and {Alarcon}, A. and {Allam}, S. and {Alves}, O. and {Amon}, A. and {Anbajagane}, D. and {Andrade-Oliveira}, F. and {Armstrong}, P. and {Avila}, S. and {Beas-Gonzalez}, J. and {Bechtol}, K. and {Becker}, M.~R. and {Bernstein}, G.~M. and {Bertin}, E. and {Blazek}, J. and {Bocquet}, S. and {Brooks}, D. and {Brout}, D. and {Burke}, D.~L. and {Camacho}, H. and {Camacho-Ciurana}, G. and {Camilleri}, R. and {Campailla}, G. and {Campos}, A. and {Carnero Rosell}, A. and {Carr}, A. and {Carretero}, J. and {Castander}, F.~J. and {Cawthon}, R. and {Chan}, K.~C. and {Chang}, C. and {Chen}, R. and {Coloma-Nadal}, J.~M. and {Conselice}, C. and {Costanzi}, M. and {Crocce}, M. and {d'Assignies}, W. and {da Costa}, L.~N. and {da Silva Pereira}, M.~E. and {Davis}, T.~M. and {De Vicente}, J. and {DePoy}, D.~L. and {DeRose}, J. and {Desai}, S. and {Diehl}, H.~T. and {Dodelson}, S. and {Doel}, P. and {Doux}, C. and {Drlica-Wagner}, A. and {Eifler}, T.~F. and {Elvin-Poole}, J. and {Everett}, S. and {Evrard}, A.~E. and {Ferrero}, I. and {Fert{\'e}}, A. and {Flaugher}, B. and {Fosalba}, P. and {de Souza}, D. Francis and {Frieman}, J. and {Galbany}, L. and {Garc{\'\i}a-Bellido}, J. and {Gatti}, M. and {Giannini}, G. and {Giles}, P. and {Glazebrook}, K. and {Gruen}, D. and {Gruendl}, R.~A. and {Gutierrez}, G. and {Harrison}, I. and {Hartley}, W.~G. and {Herner}, K. and {Hinton}, S.~R. and {Hollowood}, D.~L. and {Honscheid}, K. and {Huff}, E.~M. and {Huterer}, D. and {Jain}, B. and {James}, D.~J. and {Jarvis}, M. and {Jeffrey}, N. and {Jeltema}, T. and {Kent}, S. and {Kessler}, R. and {Kovacs}, A. and {Koyama}, K. and {Krause}, E. and {Kron}, R. and {Kuehn}, K. and {Lahav}, O. and {Lee}, J. and {Lee}, S. and {Legnani}, E. and {Li}, T.~S. and {Liddle}, A.~R. and {Lidman}, C. and {Lin}, H. and {Lin}, M. and {MacCrann}, N. and {Marshall}, J.~L. and {Mau}, S. and {McMahon}, R.~G. and {Mena-Fern{\'a}ndez}, J. and {Menanteau}, F. and {Miquel}, R. and {Mohr}, J.~J. and {Muir}, J. and {Myles}, J. and {M{\"o}ller}, A. and {Nichol}, R.~C. and {Ogando}, R.~L.~C. and {Percival}, W.~J. and {Petravick}, D. and {Pieres}, A. and {Plazas Malag{\'o}n}, A.~A. and {Popovic}, B. and {Porredon}, A. and {Prat}, J. and {Qu}, H. and {Raveri}, M. and {Rebou{\c{c}}as}, J. and {Riquelme}, W. and {Rodriguez-Monroy}, M. and {Rogozenski}, P. and {Romer}, A.~K. and {Roodman}, A. and {Rosenfeld}, R. and {Ross}, A.~J. and {Rykoff}, E.~S. and {Sako}, M. and {Samuroff}, S. and {S{\'a}nchez}, C. and {Sanchez}, E. and {Sanchez Cid}, D. and {Schutt}, T. and {Scolnic}, D. and {Sevilla-Noarbe}, I. and {Shah}, N. and {Shah}, P. and {Sheldon}, E. and {Smith}, M. and {Soares-Santos}, M. and {Suchyta}, E. and {Sullivan}, M. and {Swanson}, M.~E.~C. and {S{\'a}nchez}, B.~O. and {Tabbutt}, M. and {Tarle}, G. and {Taylor}, G. and {Thomas}, D. and {To}, C. and {Toribio San Cipriano}, L. and {Toy}, M. and {Troxel}, M.~A. and {Tucker}, D.~L. and {Vikram}, V. and {Vincenzi}, M. and {Weaverdyck}, N. and {Weller}, J. and {Whyley}, A. and {Wilkinson}, R.~D. and {Wiseman}, P. and {Yamamoto}, M. and {Yanny}, B. and {Yin}, B. and {Zhang}, Y. and {Zuntz}, J.},
        title = "{Constraints on Dynamical Dark Energy from Multiple Probes in the Full Dark Energy Survey}",
      journal = {arXiv e-prints},
         year = 2026,
        month = may,
          eid = {arXiv:2605.27221},
        pages = {arXiv:2605.27221},
          doi = {10.48550/arXiv.2605.27221},
archivePrefix = {arXiv},
       eprint = {2605.27221},
 primaryClass = {astro-ph.CO},
       adsurl = {https://ui.adsabs.harvard.edu/abs/2026arXiv260527221D}
}

@ARTICLE{KiDS2026:DE,
       author = {{Reischke}, Robert and {St{\"o}lzner}, Benjamin and {Joachimi}, Benjamin and {Wright}, Angus H. and {Asgari}, Marika and {Bilicki}, Maciej and {Elisa Chisari}, Nora and {Dvornik}, Andrej and {Georgiou}, Christos and {Giblin}, Benjamin and {Harnois-D{\'e}raps}, Joachim and {Heymans}, Catherine and {Hildebrandt}, Hendrik and {Hoekstra}, Henk and {Joudaki}, Shahab and {Kuijken}, Konrad and {Li}, Shun-Sheng and {Linke}, Laila and {Loureiro}, Arthur and {Mahony}, Constance and {Moscardini}, Lauro and {Napolitano}, Nicola R. and {Porth}, Lucas and {Radovich}, Mario and {Tr{\"o}ster}, Tilman and {von Wietersheim-Kramsta}, Maximilian and {Yan}, Ziang and {Yoon}, Mijin and {Zhang}, Yun-Hao},
        title = "{KiDS-Legacy: Constraining dark energy, neutrino mass, and curvature}",
      journal = {\aap},
         year = 2026,
        month = may,
       volume = {709},
          eid = {A82},
        pages = {A82},
          doi = {10.1051/0004-6361/202558581},
archivePrefix = {arXiv},
       eprint = {2512.11041},
 primaryClass = {astro-ph.CO},
       adsurl = {https://ui.adsabs.harvard.edu/abs/2026A&A...709A..82R}
}

@ARTICLE{Sugiyama2023:HSCY3_3x2,
       author = {{Sugiyama}, Sunao and {Miyatake}, Hironao and {More}, Surhud and {Li}, Xiangchong and {Shirasaki}, Masato and {Takada}, Masahiro and {Kobayashi}, Yosuke and {Takahashi}, Ryuichi and {Nishimichi}, Takahiro and {Nishizawa}, Atsushi J. and {Rau}, Markus M. and {Zhang}, Tianqing and {Dalal}, Roohi and {Mandelbaum}, Rachel and {Strauss}, Michael A. and {Hamana}, Takashi and {Oguri}, Masamune and {Osato}, Ken and {Kannawadi}, Arun and {Hsieh}, Bau-Ching and {Luo}, Wentao and {Armstrong}, Robert and {Bosch}, James and {Komiyama}, Yutaka and {Lupton}, Robert H. and {Lust}, Nate B. and {Miyazaki}, Satoshi and {Murayama}, Hitoshi and {Okura}, Yuki and {Price}, Paul A. and {Tait}, Philip J. and {Tanaka}, Masayuki and {Wang}, Shiang-Yu},
        title = "{Hyper Suprime-Cam Year 3 results: Cosmology from galaxy clustering and weak lensing with HSC and SDSS using the minimal bias model}",
      journal = {\prd},
         year = 2023,
        month = dec,
       volume = {108},
       number = {12},
          eid = {123521},
        pages = {123521},
          doi = {10.1103/PhysRevD.108.123521},
archivePrefix = {arXiv},
       eprint = {2304.00705},
 primaryClass = {astro-ph.CO},
       adsurl = {https://ui.adsabs.harvard.edu/abs/2023PhRvD.108l3521S}
}

@ARTICLE{Miyatake2023:HSCY3_3x2,
       author = {{Miyatake}, Hironao and {Sugiyama}, Sunao and {Takada}, Masahiro and {Nishimichi}, Takahiro and {Li}, Xiangchong and {Shirasaki}, Masato and {More}, Surhud and {Kobayashi}, Yosuke and {Nishizawa}, Atsushi J. and {Rau}, Markus M. and {Zhang}, Tianqing and {Takahashi}, Ryuichi and {Dalal}, Roohi and {Mandelbaum}, Rachel and {Strauss}, Michael A. and {Hamana}, Takashi and {Oguri}, Masamune and {Osato}, Ken and {Luo}, Wentao and {Kannawadi}, Arun and {Hsieh}, Bau-Ching and {Armstrong}, Robert and {Bosch}, James and {Komiyama}, Yutaka and {Lupton}, Robert H. and {Lust}, Nate B. and {MacArthur}, Lauren A. and {Miyazaki}, Satoshi and {Murayama}, Hitoshi and {Okura}, Yuki and {Price}, Paul A. and {Sunayama}, Tomomi and {Tait}, Philip J. and {Tanaka}, Masayuki and {Wang}, Shiang-Yu},
        title = "{Hyper Suprime-Cam Year 3 results: Cosmology from galaxy clustering and weak lensing with HSC and SDSS using the emulator based halo model}",
      journal = {\prd},
         year = 2023,
        month = dec,
       volume = {108},
       number = {12},
          eid = {123517},
        pages = {123517},
          doi = {10.1103/PhysRevD.108.123517},
archivePrefix = {arXiv},
       eprint = {2304.00704},
 primaryClass = {astro-ph.CO},
       adsurl = {https://ui.adsabs.harvard.edu/abs/2023PhRvD.108l3517M}
}

@ARTICLE{DES2026:Y6_3x2,
       author = {{DES Collaboration} and {Abbott}, T.~M.~C. and {Adamow}, M. and {Aguena}, M. and {Alarcon}, A. and {Allam}, S.~S. and {Alves}, O. and {Amon}, A. and {Anbajagane}, D. and {Andrade-Oliveira}, F. and {Avila}, S. and {Bacon}, D. and {Baxter}, E.~J. and {Beas-Gonzalez}, J. and {Bechtol}, K. and {Becker}, M.~R. and {Bernstein}, G.~M. and {Bertin}, E. and {Blazek}, J. and {Bocquet}, S. and {Brooks}, D. and {Brout}, D. and {Camacho}, H. and {Camacho-Ciurana}, G. and {Camilleri}, R. and {Campailla}, G. and {Campos}, A. and {Carnero Rosell}, A. and {Carrasco Kind}, M. and {Carretero}, J. and {Carrilho}, P. and {Castander}, F.~J. and {Cawthon}, R. and {Chang}, C. and {Choi}, A. and {Coloma-Nadal}, J.~M. and {Costanzi}, M. and {Crocce}, M. and {d'Assignies}, W. and {da Costa}, L.~N. and {da Silva Pereira}, M.~E. and {Davis}, T.~M. and {De Vicente}, J. and {DeRose}, J. and {Diehl}, H.~T. and {Dodelson}, S. and {Doel}, P. and {Doux}, C. and {Drlica-Wagner}, A. and {Eifler}, T.~F. and {Elvin-Poole}, J. and {Estrada}, J. and {Everett}, S. and {Evrard}, A.~E. and {Fang}, J. and {Farahi}, A. and {Fert{\'e}}, A. and {Flaugher}, B. and {Fosalba}, P. and {Frieman}, J. and {Garc{\'\i}a-Bellido}, J. and {Gatti}, M. and {Gaztanaga}, E. and {Giannini}, G. and {Giles}, P. and {Glazebrook}, K. and {Gorsuch}, M. and {Gruen}, D. and {Gruendl}, R.~A. and {Gschwend}, J. and {Gutierrez}, G. and {Harrison}, I. and {Hartley}, W.~G. and {Henning}, E. and {Herner}, K. and {Hinton}, S.~R. and {Hollowood}, D.~L. and {Honscheid}, K. and {Huff}, E.~M. and {Huterer}, D. and {Jain}, B. and {James}, D.~J. and {Jarvis}, M. and {Jeffrey}, N. and {Jeltema}, T. and {Kacprzak}, T. and {Kent}, S. and {Kovacs}, A. and {Krause}, E. and {Kron}, R. and {Kuehn}, K. and {Lahav}, O. and {Lee}, S. and {Legnani}, E. and {Lidman}, C. and {Lin}, H. and {MacCrann}, N. and {Manera}, M. and {Manning}, T. and {Marshall}, J.~L. and {Mau}, S. and {McCullough}, J. and {Mena-Fern{\'a}ndez}, J. and {Menanteau}, F. and {Miquel}, R. and {Mohr}, J.~J. and {Muir}, J. and {Myles}, J. and {Nichol}, R.~C. and {Nord}, B. and {O'Donnell}, J.~H. and {Ogando}, R.~L.~C. and {Palmese}, A. and {Paterno}, M. and {Peoples}, J. and {Percival}, W.~J. and {Petravick}, D. and {Pieres}, A. and {Plazas Malag{\'o}n}, A.~A. and {Porredon}, A. and {Pourtsidou}, A. and {Prat}, J. and {Preston}, C. and {Raveri}, M. and {Riquelme}, W. and {Rodriguez-Monroy}, M. and {Rogozenski}, P. and {Romer}, A.~K. and {Roodman}, A. and {Rosenfeld}, R. and {Ross}, A.~J. and {Rozo}, E. and {Rykoff}, E.~S. and {Samuroff}, S. and {S{\'a}nchez}, C. and {Sanchez}, E. and {Sanchez Cid}, D. and {Schutt}, T. and {Sevilla-Noarbe}, I. and {Sheldon}, E. and {Sherman}, N. and {Shin}, T. and {Smith}, M. and {Soares-Santos}, M. and {Suchyta}, E. and {Swanson}, M.~E.~C. and {Tabbutt}, M. and {Tarle}, G. and {Thomas}, D. and {To}, C. and {Tong}, A. and {Toribio San Cipriano}, L. and {Troxel}, M.~A. and {Tsedrik}, M. and {Tucker}, D.~L. and {Vikram}, V. and {Walker}, A.~R. and {Weaverdyck}, N. and {Wechsler}, R.~H. and {Weinberg}, D.~H. and {Weller}, J. and {Wetzell}, V. and {Whyley}, A. and {Wilkinson}, R.~D. and {Wiseman}, P. and {Wu}, H.-Y. and {Yamamoto}, M. and {Yanny}, B. and {Yin}, B. and {Zacharegkas}, G. and {Zhang}, Y. and {Zuntz}, J.},
        title = "{Dark Energy Survey Year 6 Results: Cosmological Constraints from Galaxy Clustering and Weak Lensing}",
      journal = {arXiv e-prints},
         year = 2026,
        month = jan,
          eid = {arXiv:2601.14559},
        pages = {arXiv:2601.14559},
          doi = {10.48550/arXiv.2601.14559},
archivePrefix = {arXiv},
       eprint = {2601.14559},
 primaryClass = {astro-ph.CO},
       adsurl = {https://ui.adsabs.harvard.edu/abs/2026arXiv260114559D}
}

@ARTICLE{KiDS2021:3x2,
       author = {{Heymans}, Catherine and {Tr{\"o}ster}, Tilman and {Asgari}, Marika and {Blake}, Chris and {Hildebrandt}, Hendrik and {Joachimi}, Benjamin and {Kuijken}, Konrad and {Lin}, Chieh-An and {S{\'a}nchez}, Ariel G. and {van den Busch}, Jan Luca and {Wright}, Angus H. and {Amon}, Alexandra and {Bilicki}, Maciej and {de Jong}, Jelte and {Crocce}, Martin and {Dvornik}, Andrej and {Erben}, Thomas and {Fortuna}, Maria Cristina and {Getman}, Fedor and {Giblin}, Benjamin and {Glazebrook}, Karl and {Hoekstra}, Henk and {Joudaki}, Shahab and {Kannawadi}, Arun and {K{\"o}hlinger}, Fabian and {Lidman}, Chris and {Miller}, Lance and {Napolitano}, Nicola R. and {Parkinson}, David and {Schneider}, Peter and {Shan}, HuanYuan and {Valentijn}, Edwin A. and {Verdoes Kleijn}, Gijs and {Wolf}, Christian},
        title = "{KiDS-1000 Cosmology: Multi-probe weak gravitational lensing and spectroscopic galaxy clustering constraints}",
      journal = {\aap},
         year = 2021,
        month = feb,
       volume = {646},
          eid = {A140},
        pages = {A140},
          doi = {10.1051/0004-6361/202039063},
archivePrefix = {arXiv},
       eprint = {2007.15632},
 primaryClass = {astro-ph.CO},
       adsurl = {https://ui.adsabs.harvard.edu/abs/2021A&A...646A.140H}
}

@ARTICLE{Cai2012_overlap,
       author = {{Cai}, Yan-Chuan and {Bernstein}, Gary},
        title = "{Combining weak-lensing tomography and spectroscopic redshift surveys}",
      journal = {\mnras},
         year = 2012,
        month = may,
       volume = {422},
       number = {2},
        pages = {1045-1056},
          doi = {10.1111/j.1365-2966.2012.20676.x},
archivePrefix = {arXiv},
       eprint = {1112.4478},
 primaryClass = {astro-ph.CO},
       adsurl = {https://ui.adsabs.harvard.edu/abs/2012MNRAS.422.1045C}
}

@ARTICLE{Bernstein2011_overlap,
       author = {{Bernstein}, Gary M. and {Cai}, Yan-Chuan},
        title = "{Cosmology without cosmic variance}",
      journal = {\mnras},
         year = 2011,
        month = oct,
       volume = {416},
       number = {4},
        pages = {3009-3016},
          doi = {10.1111/j.1365-2966.2011.19249.x},
archivePrefix = {arXiv},
       eprint = {1104.3862},
 primaryClass = {astro-ph.CO},
       adsurl = {https://ui.adsabs.harvard.edu/abs/2011MNRAS.416.3009B}
}

@ARTICLE{Barreira2018_SSC,
       author = {{Barreira}, Alexandre and {Krause}, Elisabeth and {Schmidt}, Fabian},
        title = "{Complete super-sample lensing covariance in the response approach}",
      journal = {\jcap},
         year = 2018,
        month = jun,
       volume = {2018},
       number = {6},
          eid = {015},
        pages = {015},
          doi = {10.1088/1475-7516/2018/06/015},
archivePrefix = {arXiv},
       eprint = {1711.07467},
 primaryClass = {astro-ph.CO},
       adsurl = {https://ui.adsabs.harvard.edu/abs/2018JCAP...06..015B}
}

@ARTICLE{Takada2013_SSC,
       author = {{Takada}, Masahiro and {Hu}, Wayne},
        title = "{Power spectrum super-sample covariance}",
      journal = {\prd},
         year = 2013,
        month = jun,
       volume = {87},
       number = {12},
          eid = {123504},
        pages = {123504},
          doi = {10.1103/PhysRevD.87.123504},
archivePrefix = {arXiv},
       eprint = {1302.6994},
 primaryClass = {astro-ph.CO},
       adsurl = {https://ui.adsabs.harvard.edu/abs/2013PhRvD..87l3504T}
}

@ARTICLE{TFR,
       author = {{Tully}, R.~B. and {Fisher}, J.~R.},
        title = "{A new method of determining distances to galaxies.}",
      journal = {\aap},
         year = 1977,
        month = feb,
       volume = {54},
        pages = {661-673},
       adsurl = {https://ui.adsabs.harvard.edu/abs/1977A&A....54..661T}
}

@ARTICLE{Schaan2020:photoz_self_calib,
       author = {{Schaan}, Emmanuel and {Ferraro}, Simone and {Seljak}, Uros},
        title = "{Photo-z outlier self-calibration in weak lensing surveys}",
      journal = {\jcap},
         year = 2020,
        month = dec,
       volume = {2020},
       number = {12},
          eid = {001},
        pages = {001},
          doi = {10.1088/1475-7516/2020/12/001},
archivePrefix = {arXiv},
       eprint = {2007.12795},
 primaryClass = {astro-ph.CO},
       adsurl = {https://ui.adsabs.harvard.edu/abs/2020JCAP...12..001S}
}

@ARTICLE{Cochrane2017:bias,
       author = {{Cochrane}, R.~K. and {Best}, P.~N. and {Sobral}, D. and {Smail}, I. and {Wake}, D.~A. and {Stott}, J.~P. and {Geach}, J.~E.},
        title = "{The H {\ensuremath{\alpha}} luminosity-dependent clustering of star-forming galaxies from z {\ensuremath{\sim}} 0.8 to {\ensuremath{\sim}}2.2 with HiZELS}",
      journal = {\mnras},
         year = 2017,
        month = aug,
       volume = {469},
       number = {3},
        pages = {2913-2932},
          doi = {10.1093/mnras/stx957},
archivePrefix = {arXiv},
       eprint = {1704.05472},
 primaryClass = {astro-ph.GA},
       adsurl = {https://ui.adsabs.harvard.edu/abs/2017MNRAS.469.2913C}
}

@ARTICLE{Merson2019_b1z,
       author = {{Merson}, Alexander and {Smith}, Alex and {Benson}, Andrew and {Wang}, Yun and {Baugh}, Carlton},
        title = "{Linear bias forecasts for emission line cosmological surveys}",
      journal = {\mnras},
         year = 2019,
        month = jul,
       volume = {486},
       number = {4},
        pages = {5737-5765},
          doi = {10.1093/mnras/stz1204},
archivePrefix = {arXiv},
       eprint = {1903.02030},
 primaryClass = {astro-ph.CO},
       adsurl = {https://ui.adsabs.harvard.edu/abs/2019MNRAS.486.5737M}
}

@ARTICLE{Zhang2026:Limber,
       author = {{Zhang}, Junzhou and {Chang}, Chihway and {Xu}, Jiachuan and {Miranda}, Vivian and {To}, Chun-Hao and {Bowden}, Haley and {Cao}, Kaili and {Eifler}, Tim and {HLIS Cosmology PIT}, Roman},
        title = "{Accurate modeling for 3$\times$2pt analyses in Roman and Rubin: a study of model approximations}",
      journal = {arXiv e-prints},
         year = 2026,
        month = jun,
          eid = {arXiv:2606.23781},
        pages = {arXiv:2606.23781},
          doi = {10.48550/arXiv.2606.23781},
archivePrefix = {arXiv},
       eprint = {2606.23781},
 primaryClass = {astro-ph.CO},
       adsurl = {https://ui.adsabs.harvard.edu/abs/2026arXiv260623781Z}
}

@ARTICLE{Planck2018,
       author = {{Planck Collaboration} and {Aghanim}, N. and {Akrami}, Y. and {Ashdown}, M. and {Aumont}, J. and {Baccigalupi}, C. and {Ballardini}, M. and {Banday}, A.~J. and {Barreiro}, R.~B. and {Bartolo}, N. and {Basak}, S. and {Battye}, R. and {Benabed}, K. and {Bernard}, J.-P. and {Bersanelli}, M. and {Bielewicz}, P. and {Bock}, J.~J. and {Bond}, J.~R. and {Borrill}, J. and {Bouchet}, F.~R. and {Boulanger}, F. and {Bucher}, M. and {Burigana}, C. and {Butler}, R.~C. and {Calabrese}, E. and {Cardoso}, J.-F. and {Carron}, J. and {Challinor}, A. and {Chiang}, H.~C. and {Chluba}, J. and {Colombo}, L.~P.~L. and {Combet}, C. and {Contreras}, D. and {Crill}, B.~P. and {Cuttaia}, F. and {de Bernardis}, P. and {de Zotti}, G. and {Delabrouille}, J. and {Delouis}, J.-M. and {Di Valentino}, E. and {Diego}, J.~M. and {Dor{\'e}}, O. and {Douspis}, M. and {Ducout}, A. and {Dupac}, X. and {Dusini}, S. and {Efstathiou}, G. and {Elsner}, F. and {En{\ss}lin}, T.~A. and {Eriksen}, H.~K. and {Fantaye}, Y. and {Farhang}, M. and {Fergusson}, J. and {Fernandez-Cobos}, R. and {Finelli}, F. and {Forastieri}, F. and {Frailis}, M. and {Fraisse}, A.~A. and {Franceschi}, E. and {Frolov}, A. and {Galeotta}, S. and {Galli}, S. and {Ganga}, K. and {G{\'e}nova-Santos}, R.~T. and {Gerbino}, M. and {Ghosh}, T. and {Gonz{\'a}lez-Nuevo}, J. and {G{\'o}rski}, K.~M. and {Gratton}, S. and {Gruppuso}, A. and {Gudmundsson}, J.~E. and {Hamann}, J. and {Handley}, W. and {Hansen}, F.~K. and {Herranz}, D. and {Hildebrandt}, S.~R. and {Hivon}, E. and {Huang}, Z. and {Jaffe}, A.~H. and {Jones}, W.~C. and {Karakci}, A. and {Keih{\"a}nen}, E. and {Keskitalo}, R. and {Kiiveri}, K. and {Kim}, J. and {Kisner}, T.~S. and {Knox}, L. and {Krachmalnicoff}, N. and {Kunz}, M. and {Kurki-Suonio}, H. and {Lagache}, G. and {Lamarre}, J.-M. and {Lasenby}, A. and {Lattanzi}, M. and {Lawrence}, C.~R. and {Le Jeune}, M. and {Lemos}, P. and {Lesgourgues}, J. and {Levrier}, F. and {Lewis}, A. and {Liguori}, M. and {Lilje}, P.~B. and {Lilley}, M. and {Lindholm}, V. and {L{\'o}pez-Caniego}, M. and {Lubin}, P.~M. and {Ma}, Y.-Z. and {Mac{\'\i}as-P{\'e}rez}, J.~F. and {Maggio}, G. and {Maino}, D. and {Mandolesi}, N. and {Mangilli}, A. and {Marcos-Caballero}, A. and {Maris}, M. and {Martin}, P.~G. and {Martinelli}, M. and {Mart{\'\i}nez-Gonz{\'a}lez}, E. and {Matarrese}, S. and {Mauri}, N. and {McEwen}, J.~D. and {Meinhold}, P.~R. and {Melchiorri}, A. and {Mennella}, A. and {Migliaccio}, M. and {Millea}, M. and {Mitra}, S. and {Miville-Desch{\^e}nes}, M.-A. and {Molinari}, D. and {Montier}, L. and {Morgante}, G. and {Moss}, A. and {Natoli}, P. and {N{\o}rgaard-Nielsen}, H.~U. and {Pagano}, L. and {Paoletti}, D. and {Partridge}, B. and {Patanchon}, G. and {Peiris}, H.~V. and {Perrotta}, F. and {Pettorino}, V. and {Piacentini}, F. and {Polastri}, L. and {Polenta}, G. and {Puget}, J.-L. and {Rachen}, J.~P. and {Reinecke}, M. and {Remazeilles}, M. and {Renzi}, A. and {Rocha}, G. and {Rosset}, C. and {Roudier}, G. and {Rubi{\~n}o-Mart{\'\i}n}, J.~A. and {Ruiz-Granados}, B. and {Salvati}, L. and {Sandri}, M. and {Savelainen}, M. and {Scott}, D. and {Shellard}, E.~P.~S. and {Sirignano}, C. and {Sirri}, G. and {Spencer}, L.~D. and {Sunyaev}, R. and {Suur-Uski}, A.-S. and {Tauber}, J.~A. and {Tavagnacco}, D. and {Tenti}, M. and {Toffolatti}, L. and {Tomasi}, M. and {Trombetti}, T. and {Valenziano}, L. and {Valiviita}, J. and {Van Tent}, B. and {Vibert}, L. and {Vielva}, P. and {Villa}, F. and {Vittorio}, N. and {Wandelt}, B.~D. and {Wehus}, I.~K. and {White}, M. and {White}, S.~D.~M. and {Zacchei}, A. and {Zonca}, A.},
        title = "{Planck 2018 results. VI. Cosmological parameters}",
      journal = {\aap},
         year = 2020,
        month = sep,
       volume = {641},
          eid = {A6},
        pages = {A6},
          doi = {10.1051/0004-6361/201833910},
archivePrefix = {arXiv},
       eprint = {1807.06209},
 primaryClass = {astro-ph.CO},
       adsurl = {https://ui.adsabs.harvard.edu/abs/2020A&A...641A...6P}
}

@ARTICLE{DESY5_SN,
       author = {{DES Collaboration} and {Abbott}, T.~M.~C. and {Acevedo}, M. and {Aguena}, M. and {Alarcon}, A. and {Allam}, S. and {Alves}, O. and {Amon}, A. and {Andrade-Oliveira}, F. and {Annis}, J. and {Armstrong}, P. and {Asorey}, J. and {Avila}, S. and {Bacon}, D. and {Bassett}, B.~A. and {Bechtol}, K. and {Bernardinelli}, P.~H. and {Bernstein}, G.~M. and {Bertin}, E. and {Blazek}, J. and {Bocquet}, S. and {Brooks}, D. and {Brout}, D. and {Buckley-Geer}, E. and {Burke}, D.~L. and {Camacho}, H. and {Camilleri}, R. and {Campos}, A. and {Carnero Rosell}, A. and {Carollo}, D. and {Carr}, A. and {Carretero}, J. and {Castander}, F.~J. and {Cawthon}, R. and {Chang}, C. and {Chen}, R. and {Choi}, A. and {Conselice}, C. and {Costanzi}, M. and {da Costa}, L.~N. and {Crocce}, M. and {Davis}, T.~M. and {DePoy}, D.~L. and {Desai}, S. and {Diehl}, H.~T. and {Dixon}, M. and {Dodelson}, S. and {Doel}, P. and {Doux}, C. and {Drlica-Wagner}, A. and {Elvin-Poole}, J. and {Everett}, S. and {Ferrero}, I. and {Fert{\'e}}, A. and {Flaugher}, B. and {Foley}, R.~J. and {Fosalba}, P. and {Friedel}, D. and {Frieman}, J. and {Frohmaier}, C. and {Galbany}, L. and {Garc{\'\i}a-Bellido}, J. and {Gatti}, M. and {Gaztanaga}, E. and {Giannini}, G. and {Glazebrook}, K. and {Graur}, O. and {Gruen}, D. and {Gruendl}, R.~A. and {Gutierrez}, G. and {Hartley}, W.~G. and {Herner}, K. and {Hinton}, S.~R. and {Hollowood}, D.~L. and {Honscheid}, K. and {Huterer}, D. and {Jain}, B. and {James}, D.~J. and {Jeffrey}, N. and {Kasai}, E. and {Kelsey}, L. and {Kent}, S. and {Kessler}, R. and {Kim}, A.~G. and {Kirshner}, R.~P. and {Kovacs}, E. and {Kuehn}, K. and {Lahav}, O. and {Lee}, J. and {Lee}, S. and {Lewis}, G.~F. and {Li}, T.~S. and {Lidman}, C. and {Lin}, H. and {Malik}, U. and {Marshall}, J.~L. and {Martini}, P. and {Mena-Fern{\'a}ndez}, J. and {Menanteau}, F. and {Miquel}, R. and {Mohr}, J.~J. and {Mould}, J. and {Muir}, J. and {M{\"o}ller}, A. and {Neilsen}, E. and {Nichol}, R.~C. and {Nugent}, P. and {Ogando}, R.~L.~C. and {Palmese}, A. and {Pan}, Y.-C. and {Paterno}, M. and {Percival}, W.~J. and {Pereira}, M.~E.~S. and {Pieres}, A. and {Malag{\'o}n}, A.~A. Plazas and {Popovic}, B. and {Porredon}, A. and {Prat}, J. and {Qu}, H. and {Raveri}, M. and {Rodr{\'\i}guez-Monroy}, M. and {Romer}, A.~K. and {Roodman}, A. and {Rose}, B. and {Sako}, M. and {Sanchez}, E. and {Sanchez Cid}, D. and {Schubnell}, M. and {Scolnic}, D. and {Sevilla-Noarbe}, I. and {Shah}, P. and {Smith}, J. Allyn. and {Smith}, M. and {Soares-Santos}, M. and {Suchyta}, E. and {Sullivan}, M. and {Suntzeff}, N. and {Swanson}, M.~E.~C. and {S{\'a}nchez}, B.~O. and {Tarle}, G. and {Taylor}, G. and {Thomas}, D. and {To}, C. and {Toy}, M. and {Troxel}, M.~A. and {Tucker}, B.~E. and {Tucker}, D.~L. and {Uddin}, S.~A. and {Vincenzi}, M. and {Walker}, A.~R. and {Weaverdyck}, N. and {Wechsler}, R.~H. and {Weller}, J. and {Wester}, W. and {Wiseman}, P. and {Yamamoto}, M. and {Yuan}, F. and {Zhang}, B. and {Zhang}, Y.},
        title = "{The Dark Energy Survey: Cosmology Results with {\ensuremath{\sim}}1500 New High-redshift Type Ia Supernovae Using the Full 5 yr Data Set}",
      journal = {\apjl},
         year = 2024,
        month = sep,
       volume = {973},
       number = {1},
          eid = {L14},
        pages = {L14},
          doi = {10.3847/2041-8213/ad6f9f},
archivePrefix = {arXiv},
       eprint = {2401.02929},
 primaryClass = {astro-ph.CO},
       adsurl = {https://ui.adsabs.harvard.edu/abs/2024ApJ...973L..14D}
}

@ARTICLE{SDSS_BAO,
       author = {{Alam}, Shadab and {Aubert}, Marie and {Avila}, Santiago and {Balland}, Christophe and {Bautista}, Julian E. and {Bershady}, Matthew A. and {Bizyaev}, Dmitry and {Blanton}, Michael R. and {Bolton}, Adam S. and {Bovy}, Jo and {Brinkmann}, Jonathan and {Brownstein}, Joel R. and {Burtin}, Etienne and {Chabanier}, Sol{\`e}ne and {Chapman}, Michael J. and {Choi}, Peter Doohyun and {Chuang}, Chia-Hsun and {Comparat}, Johan and {Cousinou}, Marie-Claude and {Cuceu}, Andrei and {Dawson}, Kyle S. and {de la Torre}, Sylvain and {de Mattia}, Arnaud and {Agathe}, Victoria de Sainte and {des Bourboux}, H{\'e}lion du Mas and {Escoffier}, Stephanie and {Etourneau}, Thomas and {Farr}, James and {Font-Ribera}, Andreu and {Frinchaboy}, Peter M. and {Fromenteau}, Sebastien and {Gil-Mar{\'\i}n}, H{\'e}ctor and {Le Goff}, Jean-Marc and {Gonzalez-Morales}, Alma X. and {Gonzalez-Perez}, Violeta and {Grabowski}, Kathleen and {Guy}, Julien and {Hawken}, Adam J. and {Hou}, Jiamin and {Kong}, Hui and {Parker}, James and {Klaene}, Mark and {Kneib}, Jean-Paul and {Lin}, Sicheng and {Long}, Daniel and {Lyke}, Brad W. and {de la Macorra}, Axel and {Martini}, Paul and {Masters}, Karen and {Mohammad}, Faizan G. and {Moon}, Jeongin and {Mueller}, Eva-Maria and {Mu{\~n}oz-Guti{\'e}rrez}, Andrea and {Myers}, Adam D. and {Nadathur}, Seshadri and {Neveux}, Richard and {Newman}, Jeffrey A. and {Noterdaeme}, Pasquier and {Oravetz}, Audrey and {Oravetz}, Daniel and {Palanque-Delabrouille}, Nathalie and {Pan}, Kaike and {Paviot}, Romain and {Percival}, Will J. and {P{\'e}rez-R{\`a}fols}, Ignasi and {Petitjean}, Patrick and {Pieri}, Matthew M. and {Prakash}, Abhishek and {Raichoor}, Anand and {Ravoux}, Corentin and {Rezaie}, Mehdi and {Rich}, James and {Ross}, Ashley J. and {Rossi}, Graziano and {Ruggeri}, Rossana and {Ruhlmann-Kleider}, Vanina and {S{\'a}nchez}, Ariel G. and {S{\'a}nchez}, F. Javier and {S{\'a}nchez-Gallego}, Jos{\'e} R. and {Sayres}, Conor and {Schneider}, Donald P. and {Seo}, Hee-Jong and {Shafieloo}, Arman and {Slosar}, An{\v{z}}e and {Smith}, Alex and {Stermer}, Julianna and {Tamone}, Amelie and {Tinker}, Jeremy L. and {Tojeiro}, Rita and {Vargas-Maga{\~n}a}, Mariana and {Variu}, Andrei and {Wang}, Yuting and {Weaver}, Benjamin A. and {Weijmans}, Anne-Marie and {Y{\`e}che}, Christophe and {Zarrouk}, Pauline and {Zhao}, Cheng and {Zhao}, Gong-Bo and {Zheng}, Zheng},
        title = "{Completed SDSS-IV extended Baryon Oscillation Spectroscopic Survey: Cosmological implications from two decades of spectroscopic surveys at the Apache Point Observatory}",
      journal = {\prd},
         year = 2021,
        month = apr,
       volume = {103},
       number = {8},
          eid = {083533},
        pages = {083533},
          doi = {10.1103/PhysRevD.103.083533},
archivePrefix = {arXiv},
       eprint = {2007.08991},
 primaryClass = {astro-ph.CO},
       adsurl = {https://ui.adsabs.harvard.edu/abs/2021PhRvD.103h3533A}
}

@ARTICLE{Chevallier_Polarski_2001_CPL,
       author = {{Chevallier}, Michel and {Polarski}, David},
        title = "{Accelerating Universes with Scaling Dark Matter}",
      journal = {International Journal of Modern Physics D},
         year = 2001,
        month = jan,
       volume = {10},
       number = {2},
        pages = {213-223},
          doi = {10.1142/S0218271801000822},
archivePrefix = {arXiv},
       eprint = {gr-qc/0009008},
 primaryClass = {gr-qc},
       adsurl = {https://ui.adsabs.harvard.edu/abs/2001IJMPD..10..213C}
}

@ARTICLE{Linder_2003_CPL,
       author = {{Linder}, Eric V.},
        title = "{Exploring the Expansion History of the Universe}",
      journal = {\prl},
         year = 2003,
        month = mar,
       volume = {90},
       number = {9},
          eid = {091301},
        pages = {091301},
          doi = {10.1103/PhysRevLett.90.091301},
archivePrefix = {arXiv},
       eprint = {astro-ph/0208512},
 primaryClass = {astro-ph},
       adsurl = {https://ui.adsabs.harvard.edu/abs/2003PhRvL..90i1301L}
}

@ARTICLE{Corte_Liddle_2024,
       author = {{Cort{\^e}s}, Marina and {Liddle}, Andrew R.},
        title = "{Interpreting DESI's evidence for evolving dark energy}",
      journal = {\jcap},
         year = 2024,
        month = dec,
       volume = {2024},
       number = {12},
          eid = {007},
        pages = {007},
          doi = {10.1088/1475-7516/2024/12/007},
archivePrefix = {arXiv},
       eprint = {2404.08056},
 primaryClass = {astro-ph.CO},
       adsurl = {https://ui.adsabs.harvard.edu/abs/2024JCAP...12..007C}
}

@ARTICLE{Toomey2026_prior,
       author = {{Toomey}, Michael W. and {Montefalcone}, Gabriele and {McDonough}, Evan and {Freese}, Katherine},
        title = "{How theory-informed priors affect DESI evidence for evolving dark energy}",
      journal = {\prd},
         year = 2026,
        month = jun,
       volume = {113},
       number = {12},
          eid = {123532},
        pages = {123532},
          doi = {10.1103/snyr-qs56},
archivePrefix = {arXiv},
       eprint = {2509.13318},
 primaryClass = {astro-ph.CO},
       adsurl = {https://ui.adsabs.harvard.edu/abs/2026PhRvD.113l3532T}
}

@ARTICLE{Colgain2026,
       author = {{{\'O} Colg{\'a}in}, Eoin and {Pourojaghi}, Saeed and {Sheikh-Jabbari}, M.~M. and {Yin}, Lu},
        title = "{How much has DESI dark energy evolved since DR1?}",
      journal = {Physics of the Dark Universe},
         year = 2026,
        month = jun,
       volume = {52},
          eid = {102268},
        pages = {102268},
          doi = {10.1016/j.dark.2026.102268},
archivePrefix = {arXiv},
       eprint = {2504.04417},
 primaryClass = {astro-ph.CO},
       adsurl = {https://ui.adsabs.harvard.edu/abs/2026PDU....5202268O}
}

@ARTICLE{Liu2024_LRG,
       author = {{Liu}, Guanlin and {Wang}, Yu and {Zhao}, Wen},
        title = "{Impact of LRG1 and LRG2 in DESI 2024 BAO data on dark energy evolution}",
      journal = {arXiv e-prints},
         year = 2024,
        month = jul,
          eid = {arXiv:2407.04385},
        pages = {arXiv:2407.04385},
          doi = {10.48550/arXiv.2407.04385},
archivePrefix = {arXiv},
       eprint = {2407.04385},
 primaryClass = {astro-ph.CO},
       adsurl = {https://ui.adsabs.harvard.edu/abs/2024arXiv240704385L}
}

@ARTICLE{Adame2025_full_shape,
       author = {{DESI Collaboration} and {Adame}, A.~G. and {Aguilar}, J. and {Ahlen}, S. and {Alam}, S. and {Alexander}, D.~M. and {Allende Prieto}, C. and {Alvarez}, M. and {Alves}, O. and {Anand}, A. and {Andrade}, U. and {Armengaud}, E. and {Avila}, S. and {Aviles}, A. and {Awan}, H. and {Bahr-Kalus}, B. and {Bailey}, S. and {Baltay}, C. and {Bault}, A. and {Behera}, J. and {BenZvi}, S. and {Beutler}, F. and {Bianchi}, D. and {Blake}, C. and {Blum}, R. and {Bonici}, M. and {Brieden}, S. and {Brodzeller}, A. and {Brooks}, D. and {Buckley-Geer}, E. and {Burtin}, E. and {Calderon}, R. and {Canning}, R. and {Carnero Rosell}, A. and {Cereskaite}, R. and {Cervantes-Cota}, J.~L. and {Chabanier}, S. and {Chaussidon}, E. and {Chaves-Montero}, J. and {Chebat}, D. and {Chen}, S. and {Chen}, X. and {Claybaugh}, T. and {Cole}, S. and {Cuceu}, A. and {Davis}, T.~M. and {Dawson}, K. and {de la Macorra}, A. and {de Mattia}, A. and {Deiosso}, N. and {Dey}, A. and {Dey}, B. and {Ding}, Z. and {Doel}, P. and {Edelstein}, J. and {Eftekharzadeh}, S. and {Eisenstein}, D.~J. and {Elbers}, W. and {Elliott}, A. and {Fagrelius}, P. and {Fanning}, K. and {Ferraro}, S. and {Ereza}, J. and {Findlay}, N. and {Flaugher}, B. and {Font-Ribera}, A. and {Forero-S{\'a}nchez}, D. and {Forero-Romero}, J.~E. and {Frenk}, C.~S. and {Garcia-Quintero}, C. and {Garrison}, L.~H. and {Gazta{\~n}aga}, E. and {Gil-Mar{\'\i}n}, H. and {Gontcho}, S. Gontcho A. and {Gonzalez-Morales}, A.~X. and {Gonzalez-Perez}, V. and {Gordon}, C. and {Green}, D. and {Gruen}, D. and {Gsponer}, R. and {Gutierrez}, G. and {Guy}, J. and {Hadzhiyska}, B. and {Hahn}, C. and {Hanif}, M.~M.~S. and {Herrera-Alcantar}, H.~K. and {Honscheid}, K. and {Howlett}, C. and {Huterer}, D. and {Ir{\v{s}}i{\v{c}}}, V. and {Ishak}, M. and {Joyce}, R. and {Juneau}, S. and {Kara{\c{c}}ayl{\i}}, N.~G. and {Kehoe}, R. and {Kent}, S. and {Kirkby}, D. and {Kong}, H. and {Koposov}, S.~E. and {Kremin}, A. and {Krolewski}, A. and {Lahav}, O. and {Lai}, Y. and {Lan}, T.-W. and {Landriau}, M. and {Lang}, D. and {Lasker}, J. and {Le Goff}, J.~M. and {Le Guillou}, L. and {Leauthaud}, A. and {Levi}, M.~E. and {Li}, T.~S. and {Lodha}, K. and {Magneville}, C. and {Manera}, M. and {Margala}, D. and {Martini}, P. and {Matthewson}, W. and {Maus}, M. and {McDonald}, P. and {Medina-Varela}, L. and {Meisner}, A. and {Mena-Fern{\'a}ndez}, J. and {Miquel}, R. and {Moon}, J. and {Moore}, S. and {Moustakas}, J. and {Mudur}, N. and {Mueller}, E. and {Mu{\~n}oz-Guti{\'e}rrez}, A. and {Myers}, A.~D. and {Nadathur}, S. and {Napolitano}, L. and {Neveux}, R. and {Newman}, J.~A. and {Nguyen}, N.~M. and {Nie}, J. and {Niz}, G. and {Noriega}, H.~E. and {Padmanabhan}, N. and {Paillas}, E. and {Palanque-Delabrouille}, N. and {Pan}, J. and {Penmetsa}, S. and {Percival}, W.~J. and {Pieri}, M.~M. and {Pinon}, M. and {Poppett}, C. and {Porredon}, A. and {Prada}, F. and {P{\'e}rez-Fern{\'a}ndez}, A. and {P{\'e}rez-R{\`a}fols}, I. and {Rabinowitz}, D. and {Raichoor}, A. and {Ram{\'\i}rez-P{\'e}rez}, C. and {Ramirez-Solano}, S. and {Rashkovetskyi}, M. and {Ravoux}, C. and {Rezaie}, M. and {Rich}, J. and {Rocher}, A. and {Rockosi}, C. and {Roe}, N.~A. and {Rosado-Marin}, A. and {Ross}, A.~J. and {Rossi}, G. and {Ruggeri}, R. and {Ruhlmann-Kleider}, V. and {Samushia}, L. and {Sanchez}, E. and {Saulder}, C. and {Schlafly}, E.~F. and {Schlegel}, D. and {Schubnell}, M. and {Seo}, H. and {Shafieloo}, A. and {Sharples}, R. and {Silber}, J. and {Slosar}, A. and {Smith}, A. and {Sprayberry}, D. and {Tan}, T. and {Tarl{\'e}}, G. and {Taylor}, P. and {Trusov}, S. and {Vaisakh}, R. and {Valcin}, D. and {Valdes}, F. and {Valogiannis}, G. and {Vargas-Maga{\~n}a}, M. and {Verde}, L. and {Walther}, M. and {Wang}, B. and {Wang}, M.~S. and {Weaver}, B.~A. and {Weaverdyck}, N. and {Wechsler}, R.~H. and {Weinberg}, D.~H. and {White}, M. and {Wilson}, M.~J. and {Yi}, L.},
        title = "{DESI 2024 VII: cosmological constraints from the full-shape modeling of clustering measurements}",
      journal = {\jcap},
         year = 2025,
        month = jul,
       volume = {2025},
       number = {7},
          eid = {028},
        pages = {028},
          doi = {10.1088/1475-7516/2025/07/028},
archivePrefix = {arXiv},
       eprint = {2411.12022},
 primaryClass = {astro-ph.CO},
       adsurl = {https://ui.adsabs.harvard.edu/abs/2025JCAP...07..028A}
}

@ARTICLE{Herold_Karwal_2026_frequentist,
       author = {{Herold}, Laura and {Karwal}, Tanvi},
        title = "{Bayesian and frequentist perspectives agree on dynamical dark energy}",
      journal = {\prd},
         year = 2026,
        month = jun,
       volume = {113},
       number = {12},
          eid = {123551},
        pages = {123551},
          doi = {10.1103/gw6q-5k5j},
archivePrefix = {arXiv},
       eprint = {2506.12004},
 primaryClass = {astro-ph.CO},
       adsurl = {https://ui.adsabs.harvard.edu/abs/2026PhRvD.113l3551H}
}

@ARTICLE{Vincenzi2025_SN_DESI,
       author = {{Vincenzi}, M. and {Kessler}, R. and {Shah}, P. and {Lee}, J. and {Davis}, T.~M. and {Scolnic}, D. and {Armstrong}, P. and {Brout}, D. and {Camilleri}, R. and {Chen}, R. and {Galbany}, L. and {Lidman}, C. and {M{\"o}ller}, A. and {Popovic}, B. and {Rose}, B. and {Sako}, M. and {S{\'a}nchez}, B.~O. and {Smith}, M. and {Sullivan}, M. and {Wiseman}, P. and {Abbott}, T.~M.~C. and {Aguena}, M. and {Allam}, S. and {Andrade-Oliveira}, F. and {Bocquet}, S. and {Brooks}, D. and {Carnero Rosell}, A. and {Carretero}, J. and {da Costa}, L.~N. and {Pereira}, M.~E.~S. and {Diehl}, H.~T. and {Doel}, P. and {Everett}, S. and {Flaugher}, B. and {Frieman}, J. and {Garc{\'\i}a-Bellido}, J. and {Gaztanaga}, E. and {Gruen}, D. and {Gruendl}, R.~A. and {Gutierrez}, G. and {Hinton}, S.~R. and {Hollowood}, D.~L. and {Honscheid}, K. and {James}, D.~J. and {Kuehn}, K. and {Lahav}, O. and {Lee}, S. and {Marshall}, J.~L. and {Mena-Fern{\'a}ndez}, J. and {Miquel}, R. and {Muir}, J. and {Myles}, J. and {Palmese}, A. and {Plazas Malag{\'o}n}, A.~A. and {Porredon}, A. and {Samuroff}, S. and {Sanchez}, E. and {Sanchez Cid}, D. and {Sevilla-Noarbe}, I. and {Suchyta}, E. and {Tarle}, G. and {To}, C. and {Tucker}, D.~L. and {Vikram}, V. and {Walker}, A.~R. and {Weaverdyck}, N. and {Weller}, J.},
        title = "{Comparing the DES-SN5YR and Pantheon+ SN cosmology analyses: investigation based on 'evolving dark energy or supernovae systematics'?}",
      journal = {\mnras},
         year = 2025,
        month = aug,
       volume = {541},
       number = {3},
        pages = {2585-2593},
          doi = {10.1093/mnras/staf943},
archivePrefix = {arXiv},
       eprint = {2501.06664},
 primaryClass = {astro-ph.CO},
       adsurl = {https://ui.adsabs.harvard.edu/abs/2025MNRAS.541.2585V}
}

@ARTICLE{Kessler2025_RomanSN,
       author = {{Kessler}, Richard and {Hounsell}, Rebekah and {Joshi}, Bhavin and {Rubin}, David and {Sako}, Masao and {Chen}, Rebecca and {Miranda}, Vivian and {Rose}, Benjamin M.},
        title = "{Cosmology Constraints from Type Ia Supernova Simulations of the Nancy Grace Roman Space Telescope Strategy Recommended by the High-Latitude Time Domain Survey Definition Committee}",
      journal = {\apj},
         year = 2025,
        month = nov,
       volume = {993},
       number = {1},
          eid = {116},
        pages = {116},
          doi = {10.3847/1538-4357/ae07c9},
archivePrefix = {arXiv},
       eprint = {2506.04402},
 primaryClass = {astro-ph.CO},
       adsurl = {https://ui.adsabs.harvard.edu/abs/2025ApJ...993..116K}
}

@ARTICLE{Mitra2026_LSSTSN,
       author = {{Mitra}, Ayan and {Kessler}, Richard and {Chen}, Rebecca C. and {Gagliano}, Alex and {Grayling}, Matthew and {More}, Surhud and {Narayan}, Gautham and {Qu}, Helen and {Raghunathan}, Srinivasan and {Malz}, Alex I. and {Lochner}, Michelle and {The LSST Dark Energy Science Collaboration}},
        title = "{A Fully Photometric Approach to Type Ia Supernova Cosmology in the LSST Era: Host-galaxy Redshifts and Supernova Classification}",
      journal = {\apj},
         year = 2026,
        month = aug,
       volume = {1006},
       number = {2},
          eid = {122},
        pages = {122},
          doi = {10.3847/1538-4357/ae7a41},
archivePrefix = {arXiv},
       eprint = {2512.06319},
 primaryClass = {astro-ph.CO},
       adsurl = {https://ui.adsabs.harvard.edu/abs/2026ApJ..1006..122M}
}

@ARTICLE{Eifler2009_cos_dep_cov,
       author = {{Eifler}, T. and {Schneider}, P. and {Hartlap}, J.},
        title = "{Dependence of cosmic shear covariances on cosmology. Impact on parameter estimation}",
      journal = {\aap},
         year = 2009,
        month = aug,
       volume = {502},
       number = {3},
        pages = {721-731},
          doi = {10.1051/0004-6361/200811276},
archivePrefix = {arXiv},
       eprint = {0810.4254},
 primaryClass = {astro-ph},
       adsurl = {https://ui.adsabs.harvard.edu/abs/2009A&A...502..721E}
}

@ARTICLE{Kodwani2019_cos_dep_cov,
       author = {{Kodwani}, Darsh and {Alonso}, David and {Ferreira}, Pedro},
        title = "{The effect on cosmological parameter estimation of a parameter dependent covariance matrix}",
      journal = {The Open Journal of Astrophysics},
         year = 2019,
        month = mar,
       volume = {2},
       number = {1},
          eid = {3},
        pages = {3},
          doi = {10.21105/astro.1811.11584},
archivePrefix = {arXiv},
       eprint = {1811.11584},
 primaryClass = {astro-ph.CO},
       adsurl = {https://ui.adsabs.harvard.edu/abs/2019OJAp....2E...3K}
}

@article{Mahalanobis,
  author = {Mahalanobis, Prasanta Chandra},
  title = {On the generalised distance in statistics},
  journal = {Proceedings of the National Institute of Sciences of India},
  volume = {2},
  pages = {49--55},
  year = {1936}
}

\end{document}